\documentclass[graphicx,11pt]{aastex} 
\usepackage{color}    

\lefthead{} \righthead{}  
\begin{document}  
 \title{Asteroids Were Born Big}  

\author{\textbf{Alessandro Morbidelli}} \affil{\small\em Observatoire de la C\^ote d'Azur\\ Boulevard de l'Observatoire\\ B.P. 4229, 06304 Nice Cedex 4, France}  

\author{\textbf{William F$.$ Bottke}} \affil{\small\em Southwest Research Institute\\  1050 Walnut St, Suite 300\\  Boulder, CO 80302~~~USA}  

\author{\textbf{David Nesvorn{\' y}}} \affil{\small\em Southwest Research Institute\\  1050 Walnut St, Suite 300\\  Boulder, CO 80302~~~USA}  

\author{\textbf{Harold F$.$ Levison}} \affil{\small\em Southwest
  Research Institute\\ 1050 Walnut St, Suite 300\\ Boulder, CO
  80302~~~USA} %\authoremail{hal@boulder.swri.edu} 

\newpage
\begin{abstract} How big were the first planetesimals? We attempt to
  answer this question by conducting coagulation simulations in which
  the planetesimals grow by mutual collisions and form larger bodies
  and planetary embryos. The size frequency distribution (SFD) of the
  initial planetesimals is considered a free parameter in these
  simulations, and we search for the one that produces at the end
  objects with a SFD that is consistent with asteroid belt
  constraints.  We find that, if the initial planetesimals were small
  (e.g. km-sized), the final SFD fails to fulfill these
  constraints. In particular, reproducing the bump observed at
  diameter $D\sim 100$~km in the current SFD of the asteroids requires
  that the {\it minimal} size of the initial planetesimals was also
  $\sim$100km. This supports the idea that planetesimals formed big,
  namely that the size of solids in the proto-planetary disk
  ``jumped'' from sub-meter scale to multi-kilometer scale, without
  passing through intermediate values. Moreover, we find evidence that
  the initial planetesimals had to have sizes ranging from 100 to {
    several 100~km, probably even 1,000~km,} and that their SFD had to
  have a slope over this interval that was similar to the one
  characterizing the current asteroids in the same size-range. This
  result sets a new constraint on planetesimal formation models and
  opens new perspectives for the investigation of the collisional
  evolution in the asteroid and Kuiper belts as well as of the
  accretion of the cores of the giant planets.  
\end{abstract}

\section{Introduction} 
\label{inro}  

The classical model for planet formation involves 3 steps.  In Step 1,
planetesimals form. Dust sediments towards the mid-plane of the
proto-planetary disk and starts to collide with each other at low
velocities.  The particles eventually stick together through
electrostatic forces, forming larger fractal aggregates (Dominik and
Tielens, 1997; Kempf et al., 1999; Wurm and Blum, 1998, 2000; Colwell
and Taylor, 1999; Blum et al., 2000). Further collisions make these
aggregates more compact, forming pebbles and larger objects. A
bottleneck for this growth mode is the so-called {\it meter-size
barrier}. The origin of this barrier is two-fold. On the one hand, the
radial drift of solid particles towards the Sun due to gas drag
reaches maximum speed for objects roughly one meter in diameter. These
meter-size boulders should fall onto the Sun from 1 AU in 100 to 1,000
years (Weidenschilling, 1977), i.e. faster than they can grow to
significantly larger sizes. On the other hand, because gas drag is
size dependent, bodies of different sizes spiral inwards at different
velocities. This leads to mutual collisions, with relative velocities
typically several tens of meters per second for bodies in the
centimeter- to meter-size range. { Moreover, in turbulent disks, even
equal-size bodies collide with non-zero velocities due to turbulent
stirring. This effect is again maximized for meter-size boulders
(Cuzzi and Weidenschilling, 2006; see also Dominik et al., 2007)}.
Current theories predict the destruction of the meter-sized objects at
these predicted speeds (Wurm et al., 2005). In the absence of a well
understood mechanism to overcome the meter-size barrier, it is usually
{\it assumed} that Nature somehow manages to produce planetesimals of
1 to 10 km in diameter, objects that are less susceptible to gas-drag
and all of its hazardous effects.

In Step 2, planetary embryos/cores form\footnote{  The embryos are
objects with Lunar to Martian masses, precursor of the terrestrial
planets, that are expected to form in the terrestrial planets region
or in the asteroid belt. In the Jovian planet region, according to the
core-accretion model (Pollack et al., 1995), the giant planet cores
are multi-Earth-mass objects that eventually lead to the birth of the
giant planets by gas accretion.}. Collisional coagulation among the
planetesimals allows the latter to agglomerate into massive bodies. In
this step gravity starts to play a fundamental role, bending the
trajectories of the colliding objects; this fact effectively increases
the collisional cross-section of the bodies by the so-called {\it
gravitational focussing factor} (Greenzweig and Lissauer, 1992). At
the beginning, if the disk is dynamically very cold (i.e. the orbits
have tiny eccentricities and inclinations), the dispersion velocity of
the planetesimals $v_{\rm rel}$ may be smaller than the escape
velocity of the planetesimals themselves. In this case, a process of
{\it runaway growth} begins, in which the relative mass growth of each
object is an increasing function of its own mass $M$, namely: 
$$ 
{1\over
M} {{{\rm d}M}\over{{\rm d}t}} \sim {{M^{1/3}}\over{v_{\rm rel}}}\ ,
$$ 
(Greenberg et al., 1978; Wetherill and Stewart, 1989). However, as
growth proceeds, the disk is dynamically heated by the scattering
action of the largest bodies. When $v_{\rm rel}$ becomes of the order
of the escape velocity from the most massive objects, the runaway
growth phase ends and the accretion proceeds in an {\it oligarchic
growth} mode, in which the relative mass growth of the largest objects
is proportional to $M^{-1/3}$ (Ida and Makino, 1993; Kokubo and Ida,
1998). The combination of runaway and oligarchic growth produces in
the inner Solar System a population of planetary embryos, with Lunar
to Martian masses (Wetherill and Stewart, 1993; Weidenschilling et
al., 1997). In the outer Solar System, beyond the so-called {\it
snowline} (Podolak and Zucker, 2004), it is generally expected that
the end result is the formation of a few super-Earth cores (Thommes et
al., 2003; Goldreich et al., 2004; Chambers, 2006) that, by accretion
of a massive gaseous atmosphere from the disk, become giant planets
(Pollack et al., 1996; Ida and Lin, 2004a,b; Alibert et al., 2004,
2005).  

In Step 3 the terrestrial planets form. The system of embryos in the
inner Solar System becomes unstable and the embryos start to collide
with each other, forming the terrestrial planets on a timescale of
several $10^7$ to $\sim 10^8$ years (Chambers and Wetherill, 1998;
Agnor et al., 1999; Chambers, 2001; Raymond et al., 2004, 2005, 2006,
2007; O'Brien et al., 2006; Kenyon and Bromley, 2006). Of all the
steps of planet formation, this is probably the one that is understood
the best, whereas Step 1 is the one that, because of the meter-size
barrier, is understood the least.

How can the meter-sized barrier be overcome? Two intriguing
possibilities come from a recent conceptual breakthrough; new models
(Johansen et al., 2007; Cuzzi et al., 2008) show that large
planetesimals can form directly from the concentration of small solid
particles in the turbulent structures of the gaseous component of the
protoplanetary disk.  Here we briefly review the models of Johansen et
al. (2007) and Cuzzi et al. (2008).

Johansen et al. (2007) showed that turbulence in the disk, either
generated by the Kelvin-Helmoltz instability (Weidenschilling, 1980;
Johansen et al., 2006) or by Magneto-Rotational Instability (MRI;
Stone et al., 2000), may help the solid particles population to
develop gravitational instabilities. Recall that turbulence generates
density fluctuations in the gas disk and that gas drag pushes
solid particles towards the maxima of the gas
density distribution. Like waves in a rough sea, these density maxima
come and go at many different locations. Thus, the concentration of
solid particles in their vicinity cannot continue for very long. The
numerical simulations of Johansen et al. (2006, 2007), however, show
that these density maxima are sufficiently long-lived (thanks also to
the inertia/feedback of the solid particles residing within the gas,
the so-called streaming instability; Youdin and Goodman, 2005) to
concentrate a large quantity of meter-size objects {  (Note
that the effect is maximized for $\sim 50$~cm objects, but we speak
of meter-size boulders for simplicity)}. Consequently, the local
density of solids can become large enough to allow the formation of a
massive planetesimal by gravitational instability.  In fact, the
simulations in Johansen et al. (2007) show the formation of a
planetesimal with 3.5 times the mass of Ceres is possible within a few
local orbital periods.

The model by Cuzzi et al. (2008) is built on the earlier result (Cuzzi
et al., 2001) that chondrule size particles are concentrated and
size-sorted in the low-vorticity regions of the disk.  In fact, Cuzzi
et al. (2008) showed that in some sporadic cases the chondrule
concentrations can become large enough to form self-gravitating
clumps. These clumps cannot become gravitationally unstable because
chondrule-sized particles are too strongly coupled to the
gas. Consequently, a sudden contraction of a chondrule clump would
cause the gas to compress, a process which is inhibited by its
internal pressure. In principle, however, these clumps might survive
in the gas disk long enough to undergo a gradual contraction,
eventually forming cohesive planetesimals roughly {  10-100km in
radius, assuming unit density}. This scenario has some advantages
over the previous one, namely that it can explain why chondrules are
the basic building blocks of chondritic planetesimals and why
chondrules appear to be size-sorted in meteorites.  Interestingly,
Alexander et al. (2008) found evidence that chondrules must form in
very dense regions that would become self-gravitating if they persist
with low relative velocity dispersion.

The models described in Johansen et al. (2007) and Cuzzi et al. (2008)
should be considered preliminary and semi-quantitative. There are a
number of open issues in each of these scenarios that are the subject
of on-going work by both teams. Moreover, there is no explicit
prediction of the size distribution of the planetesimals produced
by these mechanisms or the associated timescales needed to make a
size distribution. Both are needed, if we are to compare the results
of these models with constraints. Nevertheless, these scenarios break
the paradigm that planetesimals had to be small at the end of Step 1;
in fact, they show that large planetesimals might have formed directly
from small particles without passing through intermediate sizes. If
this is true, Step 2 was affected and visible traces should still
exist in the populations of planetesimals that still survive today: 
the asteroid belt and the Kuiper belt.

Thus, the approach that we follow in this paper is to use Step 2 to
constrain the outcome of Step 1. Our logic is as follows. We define
the {\it initial} Size Frequency Distribution (SFD) as the
planetesimal SFD that existed at the end of the planetesimal formation
phase (end of Step 1, beginning of Step 2).  We then attempt to
simulate Step 2, assuming that the initial SFD is a free parameter of
the model.  By tuning the initial SFD, we attempt to reproduce the
size distribution of the asteroid belt that existed at the end of Step
2. These simulations should allow us to glean insights into the
initial SFD of the planetesimals and, therefore, into the processes
that produced them. For instance, if we found that the size
distribution of the asteroid belt is best reproduced starting from a
population of km-size planetesimals, this would mean that the
``classical'' version of Step 1 is probably correct and that
planetesimals formed progressively by collisional coagulation. If, on
the contrary, we found that the initial SFD had to have been dominated
by large bodies, this would provide qualitative support for the new
scenarios (Johansen et al., Cuzzi et al.), namely that large
planetesimals formed directly from small objects by collective
gravitational effects. In this case, the initial SFD required by our model
would become a target function to be matched by these scenarios or by
competing ones in the future.

{  A caveat to keep in mind is that there might be an intermediate
phase between Steps 1 and 2 in which the planetesimals, initially
``fluffy'' objects with low strength, are compressed into more compact
objects by collisions and/or heat from radioactive decay. This phase,
while poorly understood, should not significantly modify the SFD
acquired in Step 1; we do not consider it here.}

The approach that we follow in this paper required us to develop
several new tools and constraints. First, in order to simulate Step 2,
we had to develop {\tt Boulder}, a statistical
coagulation/fragmentation code of the collisional accretion
process. We built this code along the lines of previous works (e.g.,
Wetherill and Stewart, 1993; {  Weidenschilling et al., 1997};
Kenyon and Luu, 1999; Kenyon and Bromley, 2001). The description of
{\tt Boulder}, as well as its validation tests are reported in the
Electronic Supplement (see http:://www.oca.eu/morby/papers/AB\_accr\_Icarus\_supp.pdf) of this
paper.

Second, the SFD that characterized the asteroid belt at the end of
Step 2 (i.e., the target function for our simulations) is not the one
currently observed. Instead, it is the SFD that the main
belt had when dynamical processes started to excite the 
asteroid orbital distribution, thus preventing any further accretion
 (Petit et al., 2002). To reconstruct the SFD
at this stage, we need to account for the collisional evolution that
occurred during and since the(se) dynamical excitation event(s)
(e.g., Bottke et al., 2005b). While no easy
task, we believe that our current models and observational constraints
are good enough to reproduce a reasonable estimate of the main belt at 
the end of the accretion phase (i.e. Step 2). Hereafter we call this
the ``reconstructed'' main belt SFD.  Below, we devote section 2 to
review the processes that the asteroid belt suffered after asteroid
accretion and discuss the properties of the reconstructed SFD.

In section 3 we assume that the initial planetesimal SFD was dominated
by km-size objects and compare the SFD obtained at the end of Step 2
with the reconstructed SFD.  In section 4, we repeat our analysis for
an extreme test case where the initial planetesimals were all 100km in
diameter. In section 5 we assume that the initial planetesimals had
sizes spanning from 100 to 500km, while in section 6 we assume that
they covered the full range of sizes from 100 km objects up to
Ceres-size bodies (~1000 km). We do this under a variety of
assumptions to show our results are generally robust. Finally, in
section 7, we summarize our results and discuss their implications for
our understanding of planetesimal accretion.

\section{Reconstructing the properties of the post-accretion asteroid
  belt}  

According to our best models (discussed below), the reconstructed main
belt {had} the following properties:
\begin {itemize}  
\item[i)] The SFD for $D > 100$~km bodies was the same as the current
main belt SFD.
\item[ii)] The SFD experienced a significant change in slope to
shallower power law values near $D \sim 100$~km.  This left a ``bump"
that can still be seen in the current main belt SFD (Fig.~1).
\item[iii)] {  The number of $D = 100$-1,000~km objects was
much larger than in the current population, probably by a factor of
100 to 1,000}
\item[iv)] The main belt included 0.01-0.1 Earth mass ($M_\oplus$)
planetary embryos.
\end {itemize}  

Below, we discuss how we obtained properties (i)--(iv) of the
reconstructed post-accretion asteroid belt (the expert reader can skip
directly to sect.~3). We are confident that the reconstructed belt is
a reasonable approximation of reality because it was worked out within
the confines of a comprehensive model that not only explains the major
properties of the observed asteroid belt but also those of the
terrestrial planets (Petit et al., 2001; O'Brien et al.,
2007). Therefore, we argue it is reasonable to use the reconstructed
belt to test predictions from planetary accretion simulations.

\subsection {The current size distribution of the main asteroid belt}  

\begin{figure}[t!]
\centerline{\includegraphics[height=7.cm]{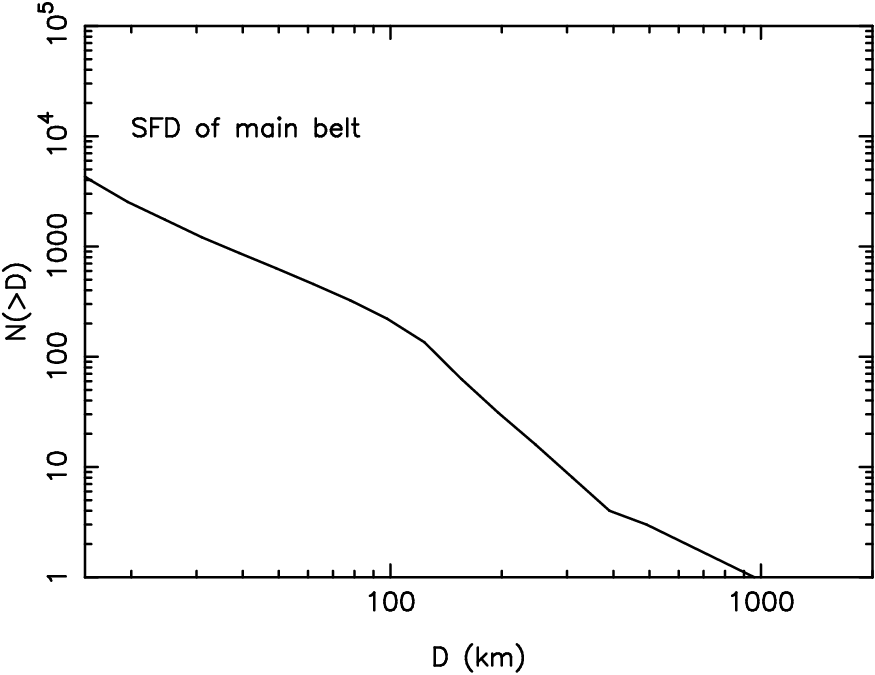}}
\vspace*{-.3cm} 
\caption{\small The size-frequency distribution (SFD)
of main belt asteroids for $D > 15$~km, assuming, for simplicity, an
albedo of $p_v = 0.092$ for all asteroids. According to Jedicke et
al. (2002), $D > 15$~km is a conservative limit for observational
completeness.} 
\label{ast_SFD} 
\end{figure}

The observed SFD of main belt asteroids is shown in
Fig.~\ref{ast_SFD}. The asteroid population throughout the main belt
is thought to be complete down to sizes of at least 15~km in diameter,
possibly even 6-10~km (Jedicke et al., 2002; Jedicke, personal
comm.). Thus, the SFD above this size threshold is the {\it real}
asteroid SFD.

\subsection {The mass deficit of the main asteroid belt }  

An estimate of the total mass of the main asteroid belt can be
obtained from the above SFD, where the largest asteroids have known
masses (Britt et al., 2002), and from an analysis of the motion of
Mars, which constrains the contribution of asteroids too small to be
observed individually (Krasinsky et al., 2002). The result is $\sim
6\times 10^{-4}$ Earth masses {  or $3.6\times 10^{24}$g}.  As we
argue below, this mass is tiny compared to the mass of solids that had
to exist in the main belt region at the time of asteroid
formation. The primordial mass of the main belt region can be
estimated by following several different lines of modeling work.

First, we consider the concept of the so-called {\it minimum mass
solar nebula} (MMSN; Weidenschilling, 1977; Hayashi, 1981).  The MMSN
implies the existence of 1-2.5 Earth masses of solid material between
2 and 3 AU.  Accordingly, this means that the main belt region is
deficient in mass by a factor 1,500-4,000. Using the same procedure,
Mars' region also appears deficient in mass, though only by a factor
of $\sim 10$. We stress that these depletion factors are actually
lower bounds because they are estimated using the concept of the {\it
minimum} mass solar nebula.  

Second, we can consider estimates of the mass of solids needed in the
main belt region for large asteroids to accrete within the time
constraints provided by meteorite data, e.g. within a few million
years (Scott, 2006).  Published results from different accretion
models, i.e. those using collisional coagulation (e.g. Wetherill,
1989) or gravitational instability (Johansen et al., 2007)
consistently find that several Earth masses of material in the main
belt region were needed to make Ceres-sized bodies within a few
My. Thus, it seems unlikely that the current asteroids were formed in
a mass-deficient environment.

Third, models of chondrule formation that assume they formed in shock
waves (Connolly and Love, 1998; Desch and Connolly, 2002; Ciesla and
Hood, 2002) require a surface density of the disk (gas plus solids) at
2.5 AU of $\sim$ 3,000~g/cm$^2$, give or take a factor of 3. Assuming
a gas/solid mass ratio of $\sim 200$ in the main belt region, this
value would correspond to a mass of solids of at least 3 Earth masses
between 2 and 3 AU.

Therefore, the available evidence is consistent with the idea that the
asteroid belt has lost more than 99.9\% of its primordial mass.  This
makes the current mass deficit in the main belt larger than a factor
of 1,000, with probable values between 2,000 and 6,000.

\subsection {Can collisions create the mass deficit found in the main
  belt region?} 
\label{collNO}  

If so much mass once existed in the primordial main belt region,
collisional evolution, dynamical removal processes, or some
combination of the two were needed to get rid of it and ultimately
produce the current main belt population.  Here we list several
arguments describing why the mass depletion was unlikely to have come
from collisional evolution of the main belt SFD.

\noindent{\it 1: Constraints from Vesta}  

The asteroid (4) Vesta is a $D = 529 \pm 10$~km differentiated body in
the inner main belt with a 25-40 km basaltic crust and one $D =
460$~km impact basin at its surface (Thomas et al., 1997).  Using a
collisional evolution model, { and assuming various size distributions
consistent with classical collisional coagulation scenarios,} Davis et
al. (1985) showed that the survival of Vesta's crust could only have
occurred {  if the asteroid belt population was only modestly larger
than it is today at the time the mean collision velocities were pumped
up to $\sim 5$km/s (i.e. the current mean impact velocity in the main
belt region; Bottke et al. 1994).} Another constraint comes from
Vesta's basin {  which formed from the impact of a $D \sim 35$~km
projectile (Thomas et al., 1997)}. The singular nature of this crater
means that Vesta, and the asteroid belt in general, could not have
been repeatedly bombarded by large (i.e. $\sim$ 30~km-sized)
impactors; otherwise, Vesta should show signs of additional basins
(Bottke et al., 2005a, 2005b; O'Brien and Greenberg, 2005). {  More
specifically, given the current collision probabilities and relative
velocities among objects of the asteroid belt, the existence of one
basin is consistent with the presence of $\sim 1000$ bodies with
$D>35$~km (i.e. the current number; see Fig.~1) over the last $\sim
4$~Gy.} {  The constraints describing Vesta¡Çs limited
collisional activity apply from the time when the asteroid belt
acquired an orbital excitation (i.e. eccentricity and inclination
distributions) comparable to the current one.}

\noindent{\it 2: Constraints from asteroid satellites}  

Collisional activity among the largest asteroids in the main belt is
also constrained by the presence of collisionally-generated satellites
(called SMATS; Durda et al., 2004).  Observations indicate that $\sim
2$\% of $D > 140$~km asteroids have SMATS (Merline et al., 2002; Durda
et al., 2004).  It was shown in Durda et al. (2004) that this fraction
is consistent, within a factor of 2--3, with the collisional activity
that the current asteroid belt population has suffered over the last
4~Gy.  If much more collisional activity had taken place, as required
by a collisional grinding scenario, one should also explain why so few
SMATS are found among the $D > 140$~km asteroids. {  Like
above, this constraint applies since the time when the asteroid belt
acquired the current orbital excitation.}

\noindent{\it 3: Constraints from meteorite shock ages}  

We also consider meteorite shock degassing ages recorded using the
$^{39}$Ar-$^{40}$Ar system. Many stony meteorite classes (e.g. L and
H-chondrites; HEDs, mesosiderites; ureilites) show evidence that the
surfaces of their parent bodies were shocked, heated, and partially
degassed by large and/or highly energetic impact events between $\sim
3.5$-4.0 Gy ago, the time of the so-called `Late Heavy Bombardment'
(e.g. Bogard, 1995; Kring and Swindle, 2008). Many meteorite classes
also show evidence for Ar-Ar degassing events on their parent body at
4.5 Gy, a time when many asteroids were experiencing metamorphism or
melting.  Curiously, the evidence for shock degassing events in the
interim between 4.1-4.4 Gy is limited, particularly when one considers
that this is the time when the asteroid belt population was expected
to be $\sim 10$ times more populous than it is today (Gomes et al.,
2005; Strom et al., 2005).  It is possible we are looking at a biased
record. For example, because shock degassing ages only record the last
resetting event that occurred on the meteorite's immediate precursor,
impacts produced by projectiles over the last 4.0 Gy may have erased
radiometric age evidence for asteroid-asteroid impacts that occurred
more than 4.0 Gy. On the other hand, many meteorite classes show clear
evidence for events that occurred 4.5 Gy.  Why did the putative
erasure events fail to eliminate these ancient Ar-Ar signatures?
While uncertainties remain, the simplest explanation is that the main
belt population experienced a minimal amount of collisional evolution
between 4.1-4.5 Gy. {  Thus, meteorites constrain the main
belt region¡Çs overall collisional activity from the time when impacts
among planetesimals became energetic enough to produce shock
degassing.}

Taken together, the above arguments imply that the asteroid belt
{  after its orbital excitation} experienced only a moderate
amount of net collisional activity over its lifetime.  Using numerical
simulations, Bottke et al. (2005a,b) found it to be roughly the
equivalent of $\sim 10$~Gy of collisional activity in the current main
belt. It is unlikely that this limited degree of collisional activity
could cause significant mass loss. In fact, the current dust
production rate of the asteroid belt is at most $10^{14}$~g~yr$^{-1}$
(i.e. assuming all IDPs come from the asteroids; Mann et al.,
1996). Thus, over a 10~Gy-equivalent of its present collisional
activity, the asteroid belt would have lost $10^{24}$~g {  only 1/3
of the current asteroid belt¡Çs mass and } a negligible amount with
respect to its inferred primordial mass of $\sim 10^{28}$~g.

If collisions {  since the orbital excitation time} cannot
explain the mass deficit of the asteroid belt, then {  the
mass either had to be lost early on, when collisions occurred at low
velocities, or by some kind of dynamical depletion mechanism. In the
first case, only small bodies could be collisionally eroded, given the
low velocities.  Even in the ``classical'' scenario, where the initial
planetesimal population was dominated by km-size bodies, it is
unlikely that more than 90\% of the initial mass could be lost in this
manner, particularly because these bodies had to accrete each other to
produce the larger asteroids observed today. We will check this
assertion in sect.~3. Accordingly, and remembering that the total mass
deficit exceeds 1,000, 99\% (or more!) of the remaining
main belt's mass had to be lost by dynamical depletion, defined here
as a} a process that excited the eccentricities of a substantial
fraction of the main belt population up to planet-crossing
values. These excited bodies would then have been rapidly eliminated
by collisions with the planets, with the Sun, or ejection from the
Solar System via a close encounter with Jupiter. We describe in
sect.~\ref{W92} the most likely process that produced this depletion
{  and its implications on the total number of objects and
size distribution of the ``post-accretion asteroid belt''.}

\subsection {Constraints provided by the main belt size distribution}  
 
The limited amount of collisional grinding that has taken place among
$D > 100$~km bodies in the asteroid belt has two additional and
profound implications. The size distribution of objects larger than
100~km could not have significantly changed since the end of accretion
(Davis et al., 1985; Durda et al., 1998; Bottke et al., 2005, 2005b,
O'Brien and Greenberg 2005). This means the observed SFD for $D >
100$~km is a primordial signature or ``fossil'' of the accretional
process. {\it This characterizes property (i) of the post-accretion
main belt population.}

Moreover, it was shown that the ``bump'' in the observed SFD at
$D=100$~km (see Fig.~\ref{ast_SFD}) {  is unlikely to be a by-product of
collisional evolution.  Bottke et al. (2005, 2005b) tested this idea
by tracking what would happen to an initial main belt SFD whose power
law slope for $D > 100$~km bodies was the same for $ D < 100$~km
bodies. Using a range of disruption scaling laws, they found they
could not grind away large numbers of $D = 50$-100 km bodies without
producing noticeable damage to the main belt SFD at larger sizes ($D =
100$-400~km) that would be readily observable today.  Other
consequences include the following. First, they found that large
numbers of $D \sim 35$~km objects would produce multiple
mega-basin-forming events on Vesta.  This is not observed.  Second,
the disruption scaling laws needed to eliminate numerous 50-100 km
asteroids would produce, over the last 3.5~Gy, far more asteroid
families from 100-200 km objects than the 20 or so current families
that are observed. Also, the ratio between the numbers of families
with progenitors larger than 100 and 200~km respectively would be a
factor of $\sim 4$ larger than observed (O'Brien and Greenberg,
2005). Finally, the asteroid belt population, and
therefore the NEO population that is sustained by the main belt, would
have decayed by more than a factor of 2 over the last 3~Gy (Davis et
al., 2002).  This is not observed in any chronology of lunar craters
(e.g. Grieve and Shoemaker, 1994).}

While Bottke et al. (2005, 2005b) could not identify the exact
power-law slope of the post-accretional SFD for $D < 100$~km objects,
their model results did suggest that it could not be steeper than what
is currently observed. {\it This sets property (ii) of the
post-accretion main belt population}. The power-law slope for $D <
100$~km could have been exceedingly shallow, with the observed SFD
derived from generations of collisional debris whose precursors were
fragments derived from break-up events among $D > 100$~km
asteroids. For this reason, in Fig.~\ref{fig-1km} and all subsequent
figures, we bracket the possible slopes of the post-accretion SFD for
$D < 100$~km by two gray lines: the slanted one representing the
current slope and the horizontal one representing an extreme case
where no bodies existed immediately below 100~km.

\subsection {The dynamical depletion of the main belt population} 
\label{W92}  

We now further explore property (iii) of the post-accretion main belt,
namely the putative dynamical depletion event that should have removed
most {  (i.e. more than 99\% in mass) of the large asteroids
as required by (a) the current total mass deficit of the asteroid belt
(a factor of at least 1,000) and (b) the relatively small mass
depletion factor that could have occurred via collisional grinding of
small bodies before the dynamical excitation event (at most a factor
of 10)}.  So far, the best model that explains the properties of the
asteroid belt is the ``indigenous embryos model'' (Wetherill, 1992;
Petit et al., 2001; O'Brien et al., 2007). Other models have been
proposed (see Petit et al. (2002), for a review) but all have problems
in reproducing at least some of the constraints, so we ignore them
here and detail briefly only the indigenous embryos model below.

According to this model (Wetherill, 1992), planetary embryos formed
not only in the terrestrial planet region but also in the asteroid
belt. The combination of their mutual perturbations and of the
dynamical action of resonances with Jupiter eventually removed them
from the asteroid belt (Chambers and Wetherill, 2001; O'Brien et al.,
2006). Before leaving the belt, however, the embryos scattered the
asteroids around them. This excited the asteroids' eccentricities and
inclinations but also forced the asteroids to random-walk in
semi-major axis. As a consequence of their mobility in semi-major
axis, many asteroids fell, at least temporarily, into resonance with
Jupiter, where their orbital eccentricities and inclinations increased
further. By this process, {  99\%} of the asteroids acquired
an eccentricity that exceeded the value characterizing the stability
boundary of the current asteroid belt {  (O'Brien et al.,
2007)}. Thus, their fate was sealed and these objects were removed
during or after the formation of the terrestrial planets.

{  In addition, about 90--95\% of the asteroids that
survived this first stage should have been removed by sweeping secular
resonances due to a sudden burst of radial migration of the giant
planets that likely triggered the so-called ``Late Heavy Bombardment"
of the Moon and the terrestrial planets (Gomes et al., 2005; Strom et
al., 2005; Minton and Malhotra, 2009). This brings
the dynamical depletion factor of the asteroid belt to a total of
$\sim$1,000. However, given the uncertainties in dynamical models and
considering that early collisional grinding among small bodies might
have removed some fraction of the initial mass, a dynamical depletion
factor of $\sim 100$ cannot be excluded. It is unlikely that the
dynamical depletion factor could be smaller than this.}

Large-scale dynamical depletion mechanisms are size independent. Thus,
a mass depletion factor of $\sim$~1,000 { (100)} implies
that the number of asteroids at the end of the accretion process had
to be, {  on average}, $\sim$1,000 {  (100)} times the
current number for all asteroid sizes.  {\it This is used to set
property (iii) of the post-accretion main belt.}

\begin{figure}[t!] 
\centerline{\includegraphics[height=7.cm]{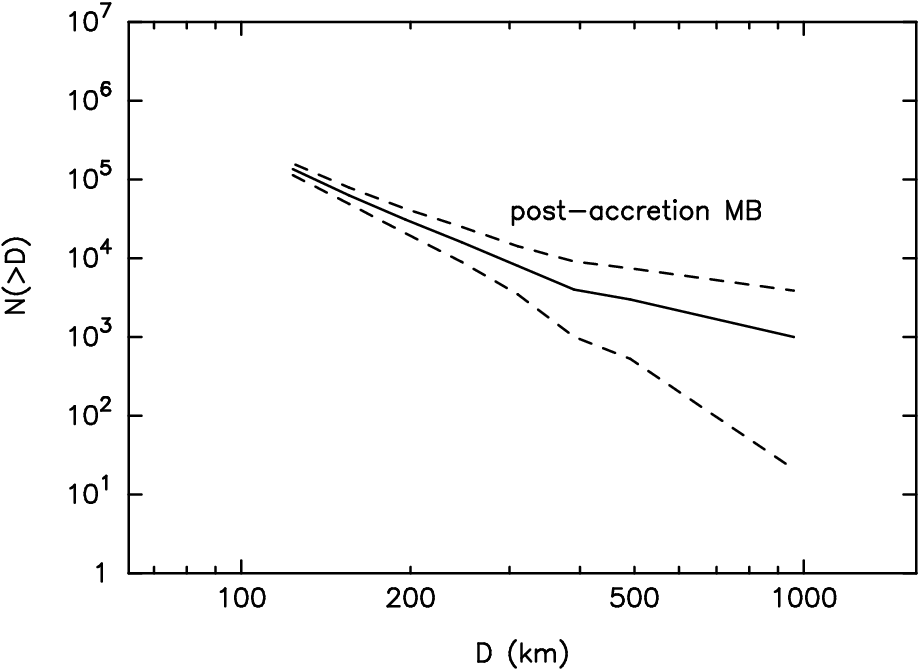}} 
%\centerline{\psfig{figure=../coag/AB_accretion/VS-DF-GD-CD/SFD-AB.ps,height=7.cm}}
%\vspace*{-.3cm} 
\caption{\small {  The size-frequency distribution (SFD) in the
    100-1,000~km range, expected for the main belt at the end of the
    accretion process (e.g. for the reconstructed belt), assuming a
    dynamical depletion factor of 1,000. The solid line is obtained
    scaling the current SFD (Fig. 1) by a factor 1,000; the dashed
    lines show the 2-$\sigma$ lower and upper bounds, given the
    current number of objects. {  The shape of these curves
      does not change (basically) with the dynamical depletion factor
      $1/p$; the curves simply shift along the $y$-axis by a quantity
      $\log((1/p)/1000)$.}}} 
\label{SFDorig} 
\end{figure}

{  At large asteroid sizes, we are affected somewhat by small number
statistics. For instance, assuming a dynamical depletion factor of
1,000, the existence of one Ceres-size body might imply the existence
of 1,000 bodies of this size, but is also consistent, at the 10\%
level, with the existence of only 100 of these bodies.

More precisely, given a population of $N$ bodies, each of which has a
probability $p$ to survive, the probability to have $M$ specimen in
the surviving population is 
\begin{equation} P=(1-p)^{N-M} p^M
N!/[M!(N-M)!]\ . 
\label{mia} 
\end{equation} 
From this, assuming that $p=10^{-3}$, one can rule out at the
2-$\sigma$ level that the population of Ceres-size bodies contained
less than 21 objects, because otherwise the odds of having one
surviving object today would be less than 2.1\%. Similarly, we can
rule out the existence of more than 3876 Ceres-size objects in the
original population, otherwise the odds of having only one Ceres today
would be smaller than 2.1\%. In an analog way, for the population of
bodies with $D>450$~km (3 objects today), the 2-$\sigma$ lower and
upper bounds on the initial population are 527 and 7441 objects,
respectively.  Fig.~\ref{SFDorig} shows the cumulative SFD of the
reconstructed asteroid belt in the 100-1,000~km range (the current SFD
scaled up by a factor 1,000; solid curve) and the 2-$\sigma$ lower and
upper bounds computed for each size as explained above (dashed
lines). The meaning of this plot is the following: consider all the
SFDs that could generate the current SFD via a random selection of 1
object every 1,000; then 95.8\% of them fall within the envelope
bounded by the dashed curves in Fig.~\ref{SFDorig}. We have checked
this result by generating these SFDs with a simple Monte-Carlo code.

We have also introduced a functional norm for these SFDs, defined as
\begin{equation} 
{\cal D}=\sum_{D_i} |\log(N'(>D_i)/N(>D_i))|\ ,
\label{D-factor} 
\end{equation} 
where $D_i$ are the size bins between 100 and 1,000~km over which the
cumulative SFD is computed (8 values), $N(>D_i)$ is the current
cumulative SFD scaled up by a factor 1,000 and $N'(>D_i)$ is a
cumulative SFD generated in the MonteCarlo code.  We have found that
95.8\% of the MonteCarlo-generated SFDs have ${\cal D}<6.14$.  By
repeating the MonteCarlo experiment with different (large) decimation
factors $p$ we have also checked that the value of ${\cal D}$ is
basically independent on $p$, while all the curves in
Fig.~\ref{SFDorig} shift along the $y$-axis proportionally to $p$ (so
that the solid line coincides with the current SFD, scaled up by a
factor $1/p$). These results will be used when testing some of our
model results in section 4 and 5.}

 The embryos-in-the-asteroid-belt-model in Wetherill (1992) does not
 only explain the depletion of the asteroid belt but also the final
 orbital excitation of eccentricities and inclinations of the
 surviving asteroids and the radial mixing of bodies of different
 taxonomic types (Petit et al., 2001). In addition, it provides a
 formidable mechanism to explain the delivery of water to the Earth
 (Morbidelli et al., 2000; Raymond et al., 2004, 2007). 

 In summary, the ``indigenous embryos'' model does a good job at
explaining the orbital and physical properties of the asteroid belt
within the larger framework of terrestrial planet accretion. To date,
it is the {\it only} model capable of doing so.  Thus, if we trust
this model, embryos of at least one lunar mass had to exist in the
primordial asteroid belt. {\it This characterizes property (iv) of the
post-accretion main belt population.} A successful accretion
simulation should not only be able to form asteroid-size bodies in the
main belt, but also a significant number of these embryos.

\section{The classical scenario: accretion from kilometer-size planetesimals}  

We start our investigation by simulating the classical version of Step
2 of the accretion process. In other words, we assume that
kilometer-size planetesimals managed to form in Step 1, despite the
meter-size barrier; the accretion of larger bodies occurrs in Step 2, by
pair-wise collisional coagulation. We simulate this second step using our
code {\tt Boulder}.  The simulations account for eccentricity $e$ and
inclination $i$ excitation due to mutual planetesimal perturbations as
well as $(e, i)$ damping due to dynamical friction, gas drag and
mutual collisions. Collisions are either accretional or disruptive
depending on the sizes of projectiles/targets and their collision
velocities.The disruption scaling law used in our simulations, defined
by the specific dispersion energy function $Q^*_D$, is the one
provided by the numerical hydro-code simulations of Benz and Asphaug
(1999) for undamaged spherical basaltic targets at impact speeds of
5km/s. See the electronic supplement for the details of the
algorithm.  {  However, in section~\ref{Zoe}, we will examine what
happens if we use a $Q^*_D$ function that allows $D < 100$ km
disruption events to occur much more easily than suggested by Benz and
Asphaug (1999), as argued in Leinhardt and
Stewart (2009) and Stewart and Leinhardt (2009).}

Here, and in all other simulations (unless otherwise specified), we
start with a total of 1.6~$M_\oplus$ in planetesimals within an
annulus between 2-3 AU.  By assuming a nominal gas/solid mass ratio of
200, this corresponds to the Minimum Mass Solar Nebula as defined in
Hayashi (1981). The bulk mass density of the planetesimals is set to 2
g/cm$^{3}$, the average value between those measured for S-type and
C-type asteroids (Britt et al., 2002). The simulations cover a
time-span of 3~My, consistent with the mean lifetime of
proto-planetary disks (Haisch et al., 2001) and hence the probable
formation timescale of Jupiter.  The initial velocity dispersion of
the planetesimals is assumed to be equal to their Hill speed (i.e. $v_{\rm
orb} [M_{\rm obj} / (3 M_{\odot})]^ {1 / 3}$, where $v_{\rm orb}$ is
the orbital speed of the object, $M_{\rm obj}$ is its mass, and
$M_{\odot}$ is the solar mass).  The lower size limit of planetesimals
tracked in our simulation is diameter $D = 0.1$~km. {  Objects
smaller than this size are removed from the simulation.  We record the
total amount of mass removed in this manner and, for brevity, refer to
it as dust.}

\begin{figure}[t!] 
\centerline{\includegraphics[height=7.cm]{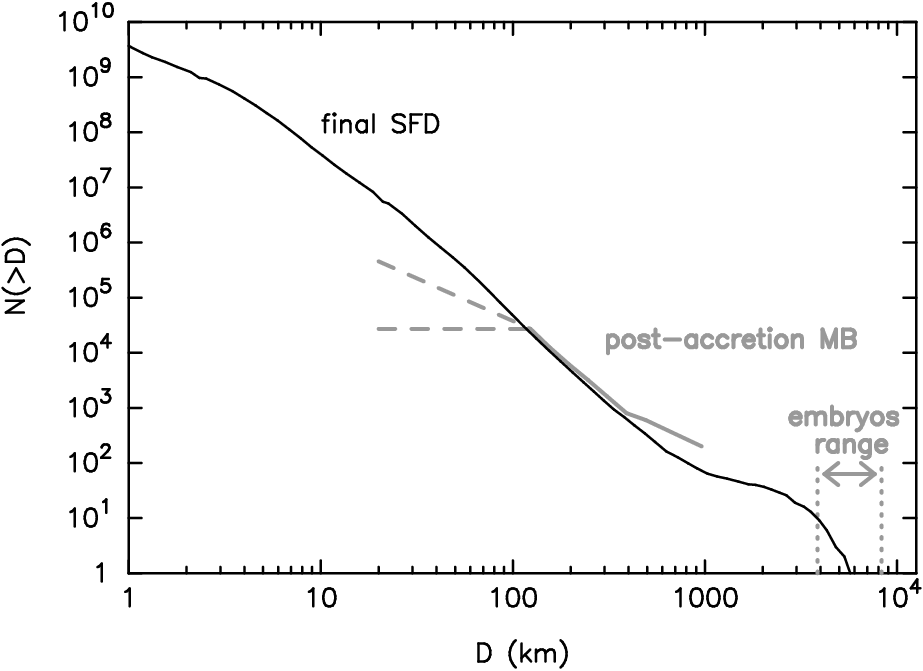}}
\vspace*{-.3cm} 
\caption{\small The gray lines show the reconstructed (i.e.,
post-accretion) main belt SFD. The solid gray curve shows the observed
main belt SFD for $100<D<1,000$~km asteroids scaled up {  200 times,
so that the number of bodies with $D>100km$ matches that produced in
the simulation}.  The dashed lines show the upper and lower bound of
the main belt power law slope in the 20-100~km range (Bottke et al.,
2005). {The upper bound corresponds to the current SFD slope}. The
vertical dotted lines show the sizes of Lunar/Martian-sized objects
for bulk density 2~g~cm$^{-3}$.  These size embryos are assumed to
have formed across the inner Solar System (Wetherill, 1992; O'Brien et
al., 2007).  The black curve shows the final SFD, starting from
$1.2\times 10^{12}$ planetesimals with $D=2$~km, at the end of the
3~My coagulation/grinding process.} 
\label{fig-1km} 
\end{figure}

The initial size of the planetesimals is assumed to be $D=2$~km.  { 
In this simulation, the total mass lost into dust by collisional
grinding is $7.45\times 10^{27}$~g, i.e. more than one Earth mass but
only 76\% of the original mass. This is consistent with our claim in
section~2 that, even starting with km-size planetesimals, low-velocity
collisions cannot deplete more than 90\% of the initial mass.}

The final SFD of the objects produced in the simulation is illustrated
by the black curve in Fig.~\ref{fig-1km}. We find this SFD does
{  not reproduce the turnover to a shallower slope that the
post accretion asteroid belt had to have at $D \sim 100$~km
(i.e. property (ii) of the reconstructed belt). Moreover,}
{  in the final SFD shown in Fig.~\ref{fig-1km}, there are
about 1.25 million bodies with $D > 35$~km. Even if we were to
magically reduce this population instantaneously by a factor $200$, in
order to reduce the number of $D>100$km bodies to the current number,
we would still have $\sim 6,200 D > 35$~km objects remaining in the
system. Recall (section~2) that $D \sim 35$~km projectiles can form
mega-basins on Vesta and that the formation of a single basin is
consistent with the existence of 1,000 of these objects in the main
belt over 4~Gy. Thus, 6,200 objects would statistically produce 6
basins; the probability that only one mega-basin would form, according
to formula (\ref{mia}), is only 1.2\%. }

{  For all these reasons, we think that this simulation produces a
result that is inconsistent with the properties of the asteroid
belt. To test whether these results are robust, we performed
additional simulations as detailed below.}

\subsection{Extending the simulation timescale}  

\begin{figure}[t!] 
\centerline{
\includegraphics[height=5.cm]{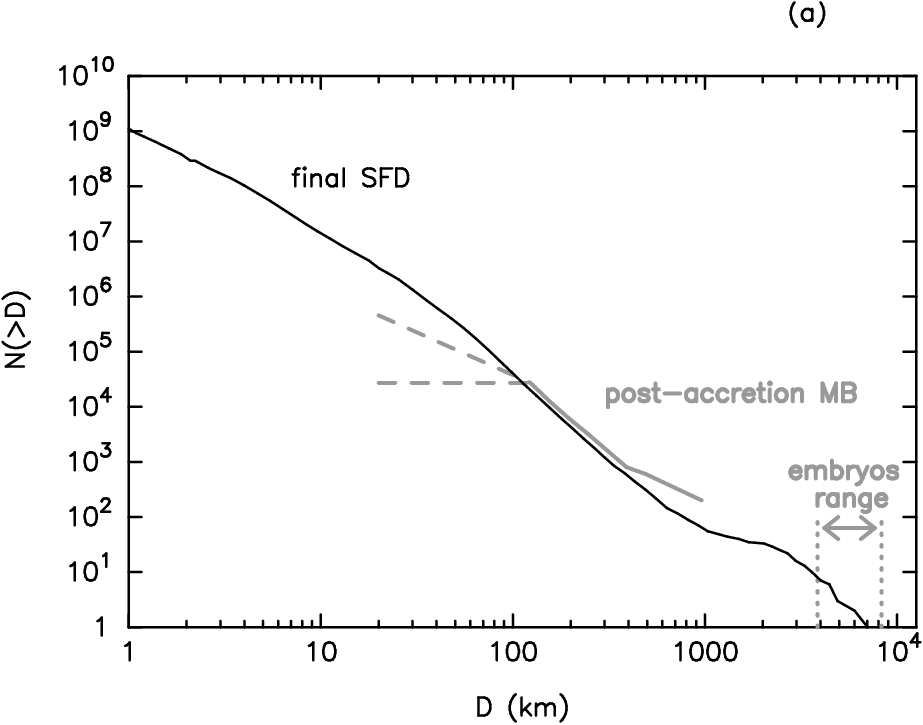} 
\includegraphics[height=5.cm]{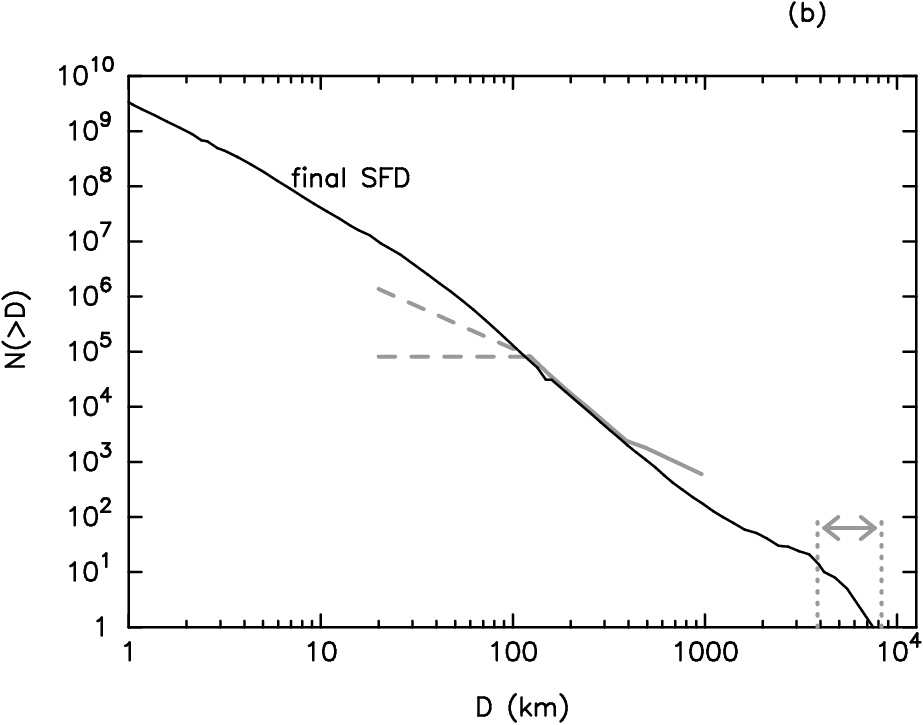}
} 
\centerline{
\includegraphics[height=5.cm]{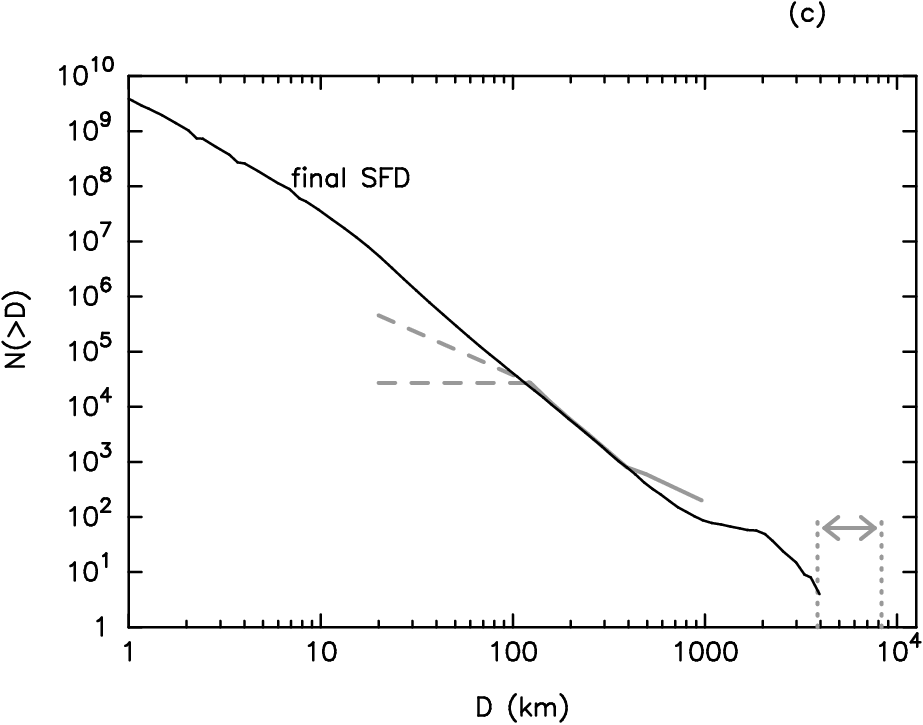} 
\includegraphics[height=5.cm]{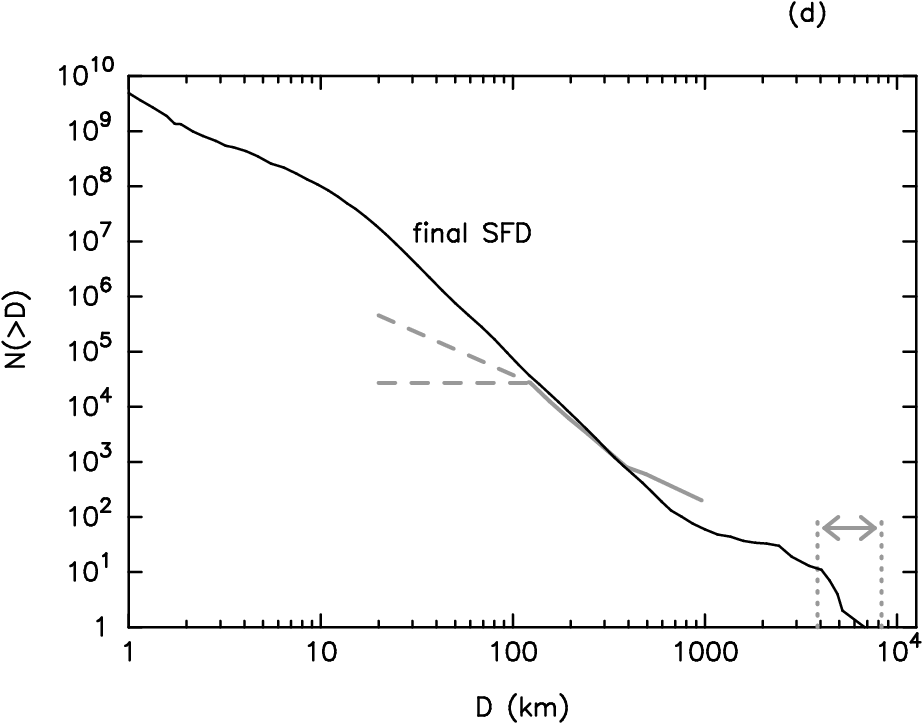}
} 
\centerline{
\includegraphics[height=5.cm]{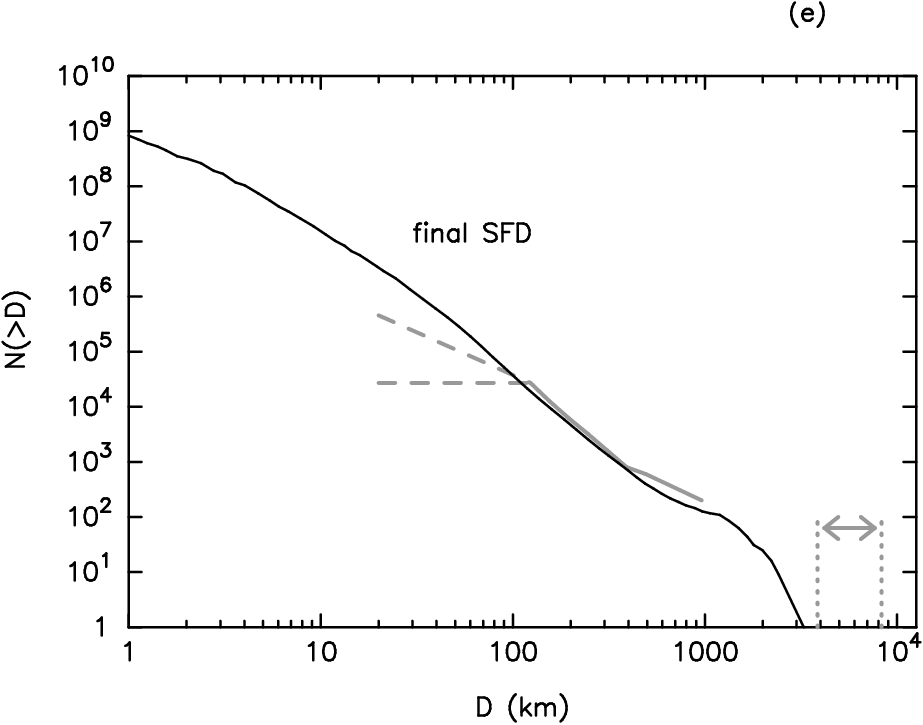}
\includegraphics[height=5.cm]{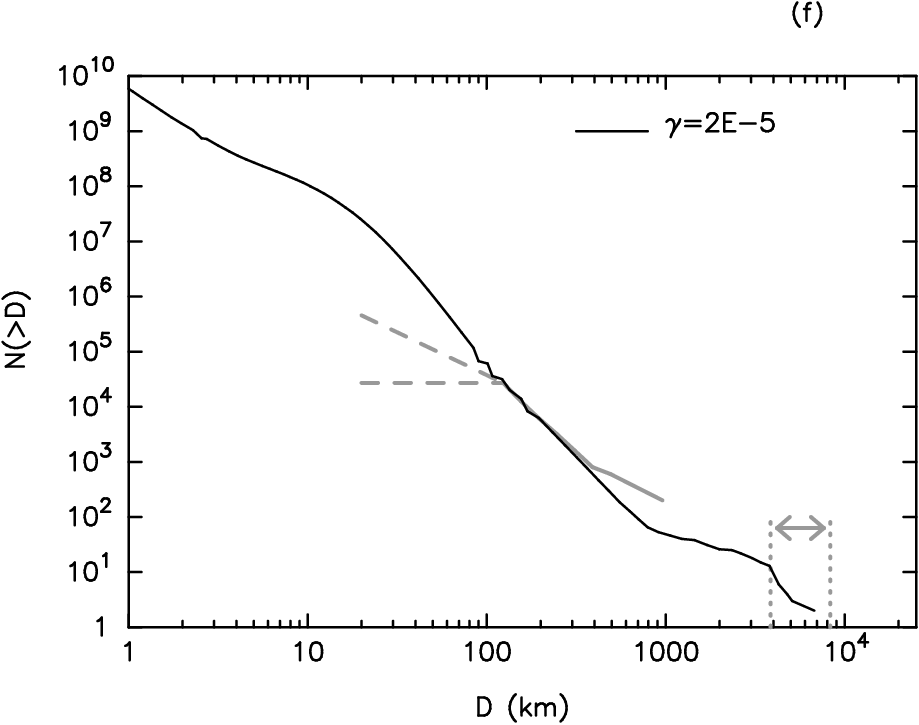}
} 
\vspace*{-.3cm} 
\caption{\footnotesize As in Fig.~\ref{fig-1km}, but for additional
simulations. (a) Continuation of the simulation of Fig.~\ref{fig-1km}
up to 10~My. (b) Starting with 5~$M_\oplus$ of $D=2$~km planetesimals
{  (here the solid gray line reproduces the current SFD
scaled up by a factor 600, instead of 200 as in all other
panels)}. (c) Starting with 1.6 ~$M_\oplus$ of $D=600$~m
planetesimals. (d) Starting with 1.6 ~$M_\oplus$ of $D=6$~km
planetesimals.  (e) Assuming that $Q^*_D$ is 1/8 of the value given in
Benz and Asphaug for ice. (f) Introducing turbulent scattering with
$\gamma=2\times 10^{-5}$ (a run with $\gamma=2\times 10^{-4}$ resulted
in basically no accretion). See text for comments on these results. }
\label{additional} 
\end{figure}

One poorly understood issue is how long the accretion phase should
last, i.e. the required length of our simulations. Thus, we continued
the simulation presented above up to 10~My. The result is illustrated
in Fig.~\ref{additional}a. {  We find that the total amount of mass
lost into dust increases only moderately, reaching at 10~My 78\% of the
mass at $t=0$}.  Also, SFD does not significantly change between
3 and 10 My.  The size of the largest embryos does grow from slightly
less than 6,000~km to about 7,000 km, mostly by agglomerating objects
smaller than a few tens of kilometers. Accretion and collisional
erosion reduce the cumulative number of $D > 1$~km objects from
$3.5\times 10^9$ to $9.3\times 10^8$.  For the size range
$80<D<$5,000~km, however, the SFD remains identical. So, the mismatch
{  with the ``bump'' observed at $D\sim 100$km does not
improve. The number of $D>35$km objects decreases slightly relative to
Fig.~\ref{fig-1km}; but the probability that only one basin is formed
on Vesta in case of an instantaneous dynamical depletion event remains
low (5\%).}

\subsection{Changing the initial mass}  

Another poorly-constrained parameter is the initial total mass of the
planetesimal population. For this reason, we tested a range of
options. Here we discuss a simulation starting with 
a system of planetesimals carrying cumulatively
5~$M_\oplus$ instead of 1.6~$M_\oplus$ as in Fig.~\ref{fig-1km}. This
total mass is of the order but slightly larger than that computed in
Weidenschilling (1977) and is the same as assumed in Wetherill (1989).
The result is shown in Fig.~\ref{additional}b. As a result of the
factor of $\sim 3$ increase in initial total mass, the final SFD is
similar to that of Fig.~\ref{fig-1km}, but scaled up by a factor of
$\sim 3$ and is nearly indistinguishable in shape for $70<D<$1,000~km
objects.

{  Thus, the turn-over of the SFD at $D\sim 100$~km is still
not reproduced.  As a consequence, there are about 2.5 million bodies
with $D>35$~km, the putative size of the basin-forming projectile on
Vesta. Invoking an instantaneous dynamical depletion event capable of
removing a factor of 600 from the population, a value needed to reduce
the the number of $D>100$km bodies to the current number, about 4,000
$D > 35$~km objects would be left in the system. Thus, about 4 basins
should have formed on Vesta; the probability that only 1 would have
formed, according to (\ref{mia}) is 6\%.}

{  For all these reasons, we think that it would be very difficult
to claim that the simulation of Fig.~\ref{additional}b is successful.
Notice that also in this case the total amount of mass lost in
collisional grinding does not exceed 86\% of the initial mass.}

\subsection{Changing the initial size of the first planetesimals}  

Fig.~\ref{additional}c and~\ref{additional}d illustrate how the
results depend on the size of the initial planetesimals. The
simulation in Fig.~\ref{additional}c starts from 1.6~$M_\oplus$ of
material in $D=600$~m planetesimals instead of $D=2$~km as in the
nominal simulation. The final SFD is indistinguishable from that of
the nominal simulation up to $D\sim$~3,000~km. Instead, there is a
deficit of larger planetary embryos. 

The simulation in Fig.~\ref{additional}d starts from the same total
mass in the form of $D=6$~km planetesimals. The final SFD has an
excess of 10-200~km objects relative to the SFD in the nominal
simulation, but the SFDs are similar in the $D = 40$-4,000~km
range. 

{  Thus, these cases can be rejected according to the same
criteria applied in Sec. 3.2 }.

\subsection{Changing the specific dispersion energy of planetesimals} 
\label{Zoe} 

{  In the previous simulations we assumed that the planetesimals
have size-dependent specific disruption energy  ($Q_D^*$) characteristic of
undamaged basalt targets being hit at several km/s (see Benz and
Asphaug, 1999). Leinhardt and Stewart (2009) have argued that the
original planetesimals might have been weak aggregates with little
strength. Moreover, Stewart and Leinhardt (2009) showed that early
planetesimals should have low $Q_D^*$ also because impact
energy couples to the target object better at low velocities.  In
these conditions, $Q_D^*$ might be more than an order of magnitude
weaker than the one that we adopted at all sizes. To test how the
results change for extremely weak material, we have re-run the
coagulation simulation starting with 1.6~$M_\oplus$ in $D=2$km
planetesimals (that of Fig.~\ref{fig-1km}), this time assuming $Q_D^*$
is one eighth of that reported by Benz and Asphaug (1999) for
competent ice struck at impact velocities of 1km/s. This is fairly
close to the value found by Leinhardt and Stewart (2009) for
strenghtless planetesimals.

 The resulting SFD is shown in Fig.~\ref{additional}e. Overall, the
 outcome is not very different from that of the nominal
 simulation. Despite of the weakness of the objects, the total mass
 lost in collisional grinding ($8.4\times 10^{27}$g) does not exceed
 90\% of the initial mass, as we argued in section 2.  Interestingly,
 though, this simulation fails to form objects more massive than our
 Moon. Thus, in conclusion, the change to a new disruption scaling law
 produces a worse fit to the constraints than before, particularly
 because constraint (iv) of the reconstructed belt (e.g. the existence
 of Lunar-to-Martian mass embryos) is not fulfilled.}

\subsection{The effect of turbulence} 
\label{turbuN}  

{  In all previous simulations we have implicitly assumed that the
gas disk in which the planetesimals evolve is laminar. Thus, the gas
can only damp the velocity dispersion of the planetesimals. In this
case, the sole mechanism enhancing the planetesimal velocity
dispersion is provided by mutual close encounters, also named {\it
viscous stirring} (Wetherill and Stewart, 1989; see section.~1.4.1 of
the electronic supplement).  In reality, the disk should be turbulent
at some level. As discussed in Cuzzi and Weidenschilling (2006), local
turbulence contributes by stirring the particles and increasing their
velocity dispersion. This effect is maximized for meter-sized
objects. In addition, however, turbulent disks show large-scale
fluctuations in gas density (Papaloizou and Nelson, 2003). The
fluctuating density maxima act as gravitational scatterers on the
planetesimals, providing an additional mechanism of excitation for the
velocity dispersion that is independent of the planetesimal masses
(Nelson et al., 2005). To distinguish this mechanism from that
discussed by Cuzzi and Weidenschilling, we call it {\it turbulent
scattering} hereafter. Ida et al. (2008) showed with simple
semi-analytical considerations that turbulent scattering can be a
bottleneck for collisional coagulation because it can move collisions
from the accretional regime to the disruptive regime. Here, we check
this result with our code.

 {\tt Boulder} accounts for turbulent scattering using the recipe
 described in Ida et al. (2008) and detailed in sect.~1.4.7 of the
 electronic supplement of this paper. In short, in the equations for
 the evolution of the velocity dispersion, there is a parameter
 $\gamma$ governing ``turbulence strength''. The effective value of
 $\gamma$ in disks that are turbulent due to the magneto-rotational
 instability is uncertain by at least an order of
 magnitude. Simulation by Laughlin et al. (2004) suggest that
 $\gamma\sim 10^{-3}$--$10^{-2}$, but values as low as $10^{-4}$
 cannot be excluded (Ida et al., 2008). The relationship between
 $\gamma$ and the more popular parameter $\alpha$ that governs the
 viscosity in the disk in the Shakura and Sunyaev (1973) description
 {has been recently investigated in details by Baruteau (2009). He
 found that $\alpha\propto \gamma^2/(h/a)^2$ where $h/a$ is the
 scale-height of the gas disk; for $h/a=3$\%, $\gamma=10^{-4}$
 corresponds to $\alpha\sim 5\times 10^{-4}$.} 

  We have re-run the coagulation simulation of Fig.~\ref{fig-1km},
  assuming $\gamma=2\times 10^{-4}$ {(which corresponds to
  $\alpha\sim 2\times 10^{-3}$ according to Baruteau's scaling).}. In
  this run we adopt an initial velocity dispersion of the
  planetesimals that is larger than that assumed in the non-turbulent
  simulations illustrated above. Recall that in all previous
  simulations the initial velocity dispersion of the objects was set
  equal to their Hill velocity. These velocities are too small for a
  turbulent disk. If we adopted them, we would get a spurious initial
  phase of fast accretion, before the velocities were fully stirred up
  by the turbulent disk. Thus, we need to start with velocity
  dispersions that represent the typical values achieved in the disk.
  More precisely, for $D = 2$~km objects, we assume initial
  eccentricities and inclinations that are the equilibrium values
  obtained by balancing the stirring effect of the turbulent disk with
  the damping effects due to gas drag and mutual collisions (Ida et
  al., 2008). We used {\tt Boulder} to estimate what these values
  should be by suppressing collisional coagulation/fragmentation and
  letting the velocity dispersion evolve from initially circular and
  co-planar orbits. We found that at equilibrium we get $e\sim 2i\sim
  2.5\times 10^{-3}$. This value is attained in about 50,000 years,
  whereas $e\sim 2i\sim 1.2\times 10^{-3}$ is attained in 5,500 years.

  In the simulation performed with this set-up, growth is fully
  aborted. The largest planetesimals produced in 3~My are just 2.5
  kilometer in diameter, whereas $9.58\times 10^{27}$g are lost in
  dust due to collisional grinding. This result is due to the fact
  that collisions become rare (because the gravitational focussing
  factor is reduced to unity by the enhanced velocity dispersion) and
  barely accretional even for ``strong'' $Q_D^*$ disruption functions
  that is used in this simulation (for basaltic targets hit at 5km/s;
  Benz and Asphaug, 1999).  We also ran a simulation where we did not
  modify the initial velocity dispersion of the planetesimals,
  although we consider this unrealistic for the reasons explained
  above. In this case there is a short initial phase of growth, as
  expected, which rapidly shuts off; the largest objects produced have
  $D=40$km. These results confirm the analysis of Ida et al. (2008);
  accretion is impossible in turbulent disks if all planetesimals are
  small.

 To investigate how weak ``turbulence strength'' should be to allow
 accretion from $D = 1$~km planetesimals, we also ran a simulation
 assuming $\gamma=2\times 10^{-5}$. In this case, we set as initial
 values $e=2i=2.5\times 10^{-4}$.  {Using the Baruteau's scaling,
 this value of $\gamma$ corresponds to $\alpha\sim 2\times 10^{-5}$,
 that is well below a minimum reasonable value in a turbulent disk};
 however, it might be acceptable for a {\it dead zone}, e.g. a region
 of the disk where the magneto-rotational instability is not at
 work. The solid curve in Fig.~\ref{additional}f shows the final SFD
 in this simulation. It now looks similar to that obtained in the
 nominal simulation of Fig.~\ref{fig-1km}, which had no turbulent
 scattering. Thus, this very low level of turbulence does not inhibit
 growth, but like the nominal simulation in Fig. 3, the resulting SFD
 is inconsistent with that of the reconstructed main belt.}

\subsection{Conclusions on the classical scenario}  

{  From the simulations illustrated in this section, we conclude
that the SFD of the initial planetesimals were not dominated by
objects with sizes the order of one kilometer. In fact, in a turbulent
disk, 1 km planetesimals would not have coagulated to form larger
bodies. In a dead zone, collisional coagulation would have produced a
final SFD that is inconsistent with the current SFD in the main
asteroid belt {  because the bump at $D\sim 100$~km is not
reproduced}; also {  we find it unlikely (at the few percent
level) that only one big basin formed on Vesta with such a SFD, even
in the case of an instantaneous dynamical depletion event of the
appropriate magnitude}.  While we were writing the final revisions of
this paper, we became aware that Weidenschilling (2009) reached the
same conclusions with similar non-turbulent simulations performed with
a different code.}

Obviously, there is an enormous parameter space left to explore, and
-strictly speaking- an infinite number of simulations would be
necessary to {\it prove} that the SFD of the reconstructed
post-accretion main belt is incompatible with the classical
collisional accretion model starting from km-size
planetesimals. Nevertheless, we believe that the 9 simulations
presented above are sufficient enough to argue that our result is
reasonably robust.

Given this conclusion, in the next sections we try to constrain which
initial planetesimal SFD would lead, at the end of Step 2, to the SFD
of the reconstructed main belt.

\section{Accretion from 100~km planetesimals}  

We start our search for the optimal initial planetesimal SFD by
assuming that all planetesimals originally had $D=100$~km. Note that
no formation model predicts that the initial planetesimals had to have
the same size. We make this assumption as a test case to probe the
signature left behind in the final SFD by the initial size of the
objects. More specifically, we attempt to satisfy property (ii), the
turnover of the size distribution at $D \sim 100$~km, assuming that
this might be the signature of the minimal size of the initial
planetesimals.

As before, our input planetesimal population carries cumulatively
1.6~$M_\oplus$. This implies that there are initially $9.4 \times
10^6$ planetesimals. The coagulation simulation covers a 3~My
time-span. {  No turbulent scattering is applied}.

\begin{figure}[t!]
\centerline{
\includegraphics[height=5.cm]{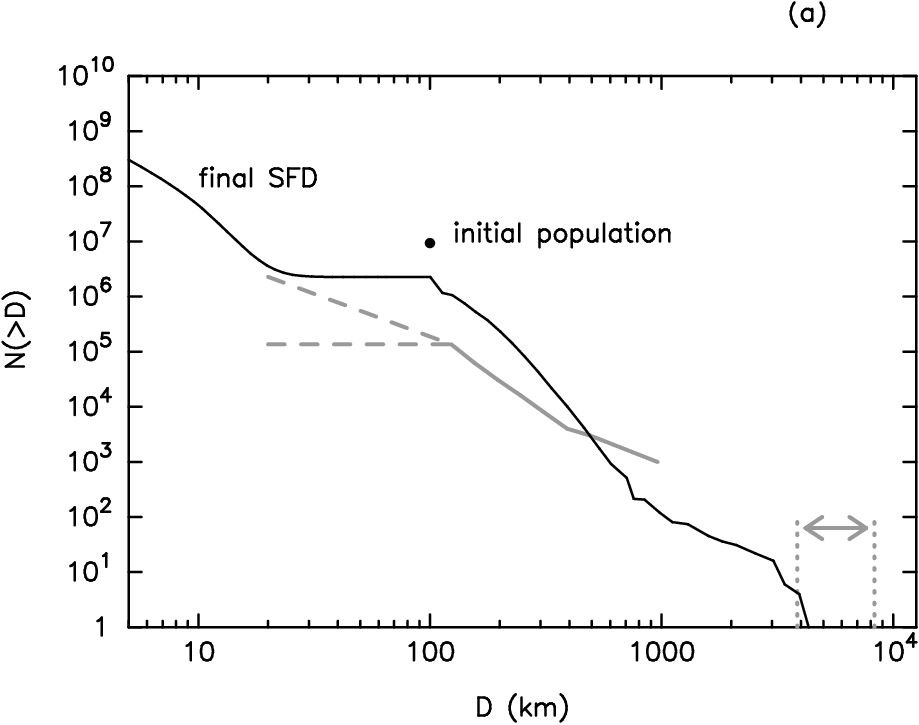}\includegraphics[height=5.cm]{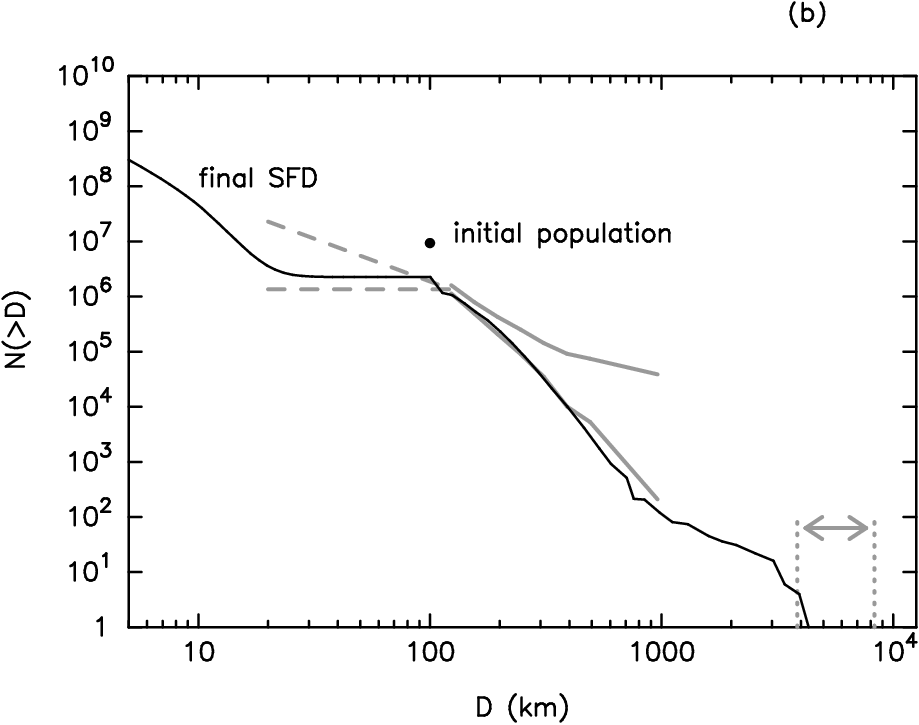}
}
\vspace*{-.3cm} 
\caption{\small {  (a): Non-turbulent simulation
starting with 1.6~$M_\oplus$ in $D=100$~km objects. The bullet shows
the initial size and total number of planetesimals. The black curve
reports the final SFD obtained after 3~My of collisional
coagulation. The grey lines sketch the reconstructed asteroid belt SFD
as in Figs.~\ref{fig-1km} and~\ref{additional}. (b): like panel (a),
but in this case the grey solid lines report the 2-$\sigma$ bounds of
the SFD of the reconstructed asteroid belt, from
Fig.~\ref{SFDorig}. Moreover, all grey lines have been moved upwards
by a factor of 10, to match the number of $D>100$km objects in the SFD
resulting from the simulation. } } 
\label{fig-100kmN} 
\end{figure}

The final SFD is shown in Fig.~\ref{fig-100kmN}a (solid curve). { 
This SFD has the same properties of that obtained by Weidenschilling
(2009) starting from $D=50$km planetesimals.} A sharp turnover of the
SFD is observed at the initial planetesimal size. {  This is in
agreement with the observed ``bump'' (i.e. property (ii) of the
reconstructed belt). However, the final SFD is much steeper than the
SFD of the current asteroid belt. Nevertheless, it would be premature
to consider this simulation unsuccessful because we showed in
Fig.~\ref{SFDorig} that the slope of the SFD of the reconstructed
asteroid belt has a large uncertainty. Thus, in
Fig.~\ref{fig-100kmN}b, we replot the final SFD against the 2-$\sigma$
bounds of the reconstructed main belt SFD. These bounds have been
taken from Fig.~\ref{SFDorig} and are ``scaled up'' by a factor of 10
so that they match the total number of $D>100$~km objects found in the
simulation. As one can see, the final SFD falls slightly out of the
lower bound of the reconstructed SFD. This means that the result is
inconsistent, at 2-$\sigma$, with the data (i.e. with the current
SFD).

Another way to check the statistical match between the simulation SFD
and the reconstructed main belt SFD is through the parameter ${\cal
D}$ defined in (\ref{D-factor}).  The SFD resulting from this
simulation has ${\cal D}=8.20$; only 0.5\% of the SFDs generated from
the current SFD in a Monte-Carlo code have ${\cal D}$ larger than this
number.  Thus, we can actually reject the result of this simulation as
inconsistent with the reconstructed main belt at nearly the 3-$\sigma$
level.}

{  Rejecting this simulation, however, is not enough to exclude the
possibility that the initial planetesimals were $\sim 100$~km in
size.} Before accepting this conclusion, we need to more extensively
explore parameter space. The simulation reported in
Fig.~\ref{fig-100kmN} is indeed simplistic {  because it did not
account for the effects of turbulence in the disk}. Recall, however,
that the works that motivated us to start with large planetesimals
(Johansen et al., 2007; Cuzzi et al., 2008) assumed (and required) a
turbulent disk, so we need to cope with turbulence effects.
Turbulence should affect our simulation in two respects: (I)
theoretical considerations (Cuzzi et al., 2008) indicate that
planetesimals should form sporadically over the lifetime of the gas
disk, in qualitative agreement with meteorite data (Scott, 2006), { 
whereas in the previous simulation we introduced all the planetesimals
at $t=0$; (II) turbulent scattering should enhance the velocity
dispersion of the planetesimals, as we have seen in
sect.~\ref{turbuN}.}. With a new suite of more sophisticated
simulations, we now attempt to circumvent our model
simplifications. We do this in steps, first addressing issue (I),
still in the framework of a laminar disk, and then (II).

To account for (I), we randomly introduce planetesimals in {\tt
Boulder} over a 2~My time-span in two different ways. In case-A, we
assume all the mass was initially in small bodies. Every time a 100-km
planetesimal is injected in the simulation, we remove an equal amount
of mass from the small bodies. In case-B, we inject equal mass
proportions of small bodies and planetesimals. This second case mimics
the possibility that planetesimal formation is regulated by the
availability of `building blocks'. Note that chondrules may be such
building blocks; they are an essential component of many meteorites
and they appear to have formed progressively over time (Scott,
2006). In both cases, we model the small body population with $D=2$~m
particles, which might be considered as tracers, representing a
population of smaller bodies of the same total mass (for instance
chondrule-size particles in the model of Cuzzi et al., or meter-size
boulders in the model of Johansen et al.). {In the previous
section, bodies of any size accreted or disrupted depending on the
impact energy relative to their specific disruption energy, consistent
with the classical scenario of planetesimal accretion.  Here, we
change our prescription. We assume that our small-bodies/tracers do
not disrupt or accrete upon mutual collisions. The rationale for this
comes from the models of Johansen et al. and Cuzzi et al. and is
twofold. First, bodies so small have difficulty sticking to one other,
so that they can not grow by binary collisions; when they form large
planetesimals, they do so thanks to their collective gravity. Second,
a large number of small bodies have to be in the disk at all times, in
order to be able to generate planetesimals over the spread of
timescales shown by meteorite data (Scott, 2006). However, the typical
relative velocities of the small bodies are not very small, because of
the effects of turbulence. Thus, either the small bodies are very
strong or, if they break, they must be rapidly regenerated by whatever
process formed them from dust grains in first place. }

\begin{figure}[h!] 
\centerline{
\includegraphics[height=5.cm]{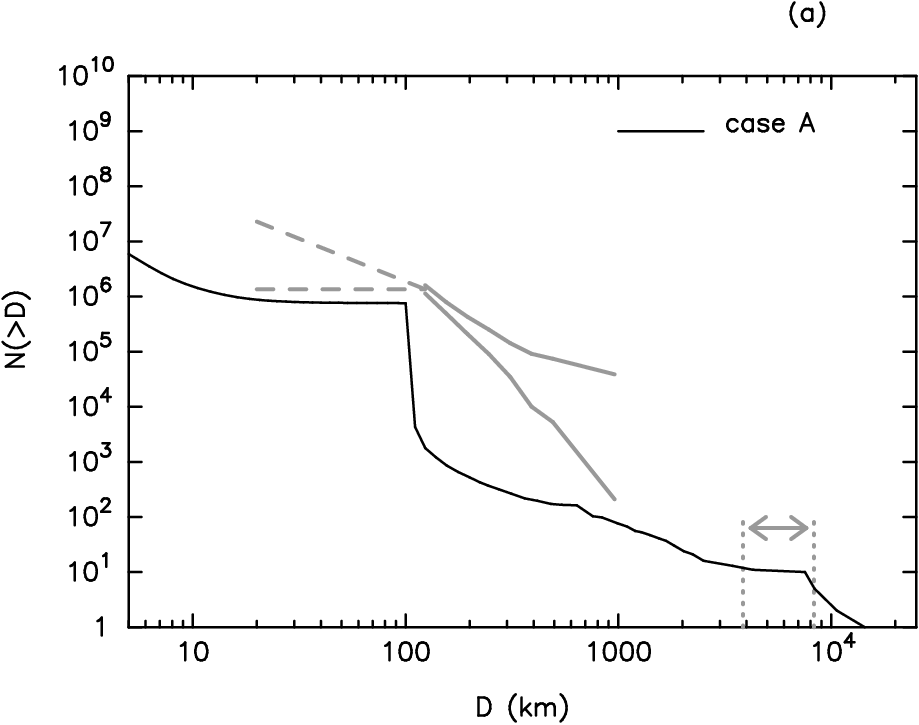} 
\includegraphics[height=5.cm]{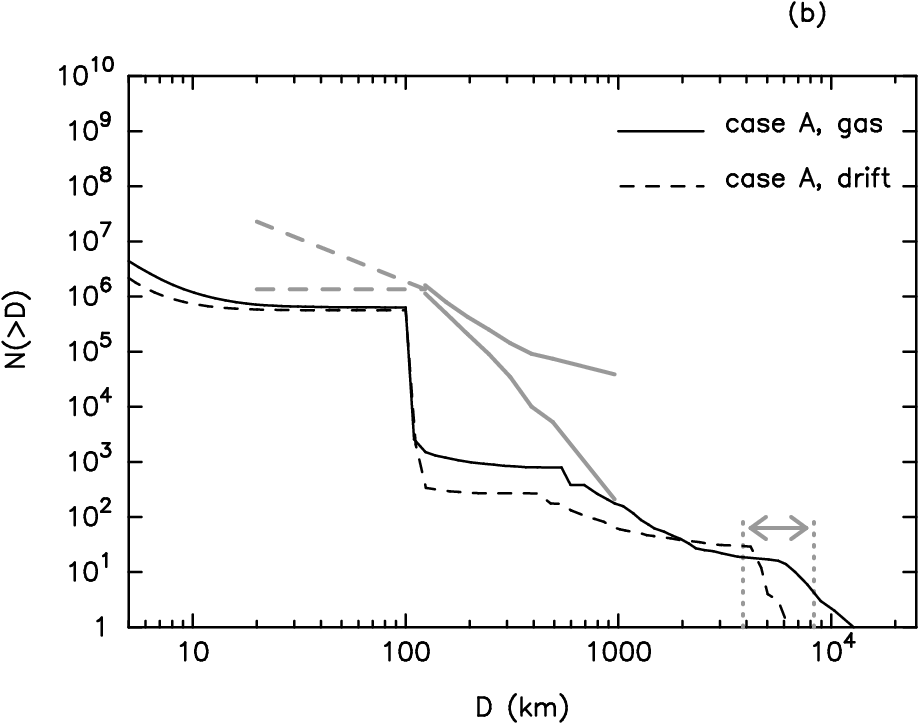}
} 
\centerline{
\includegraphics[height=5.cm]{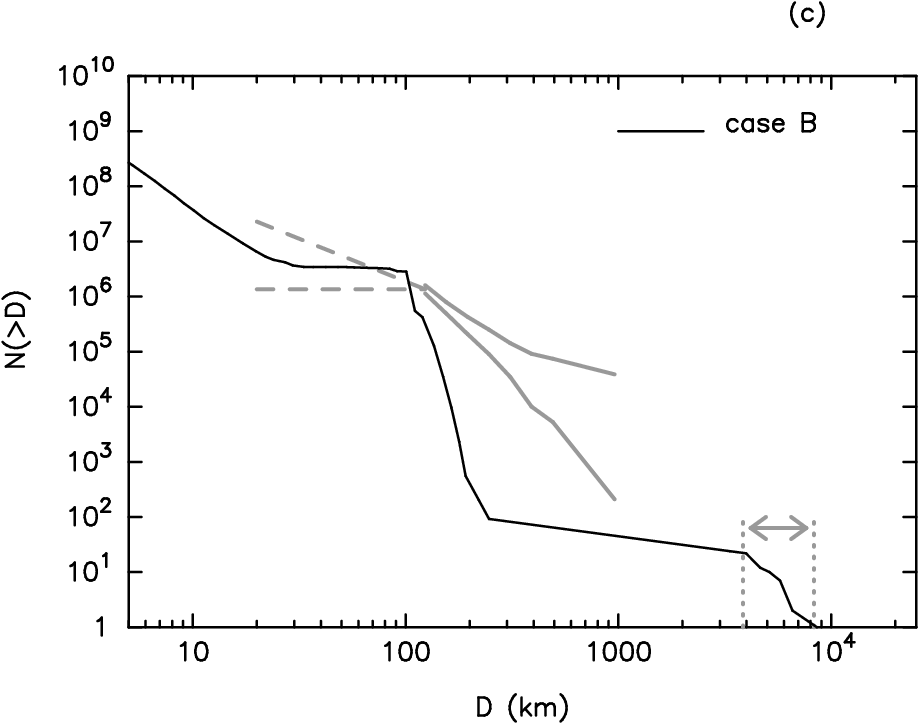} 
\includegraphics[height=5.cm]{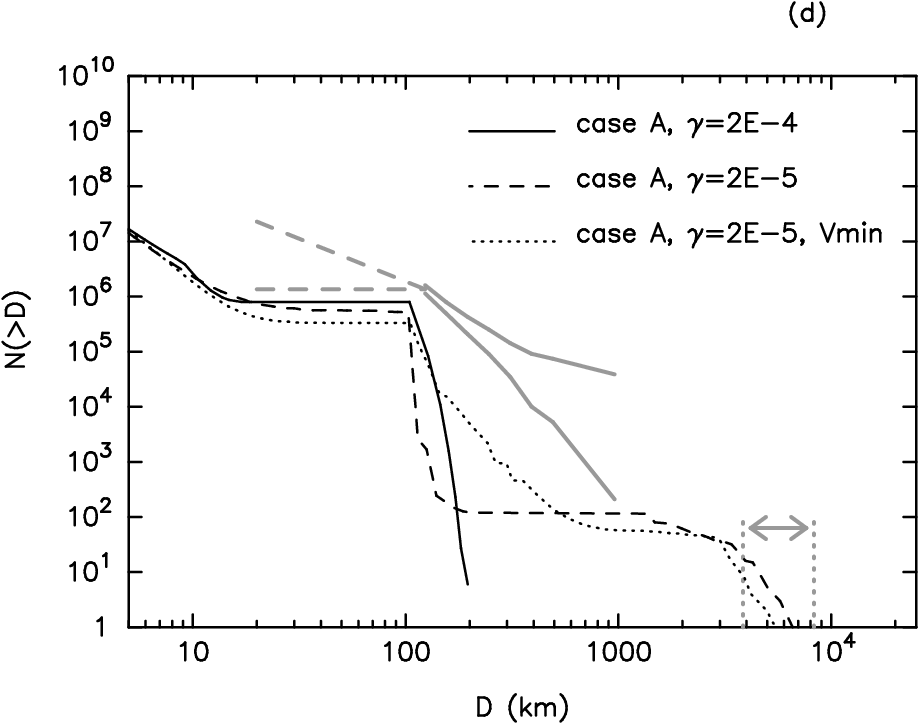}
} 
\vspace*{-.3cm} 
\caption{\footnotesize Additional coagulation simulations with 100~km
initial planetesimals. The black curves show the final SFDs; the grey
lines sketch the reconstructed asteroid belt SFD and its 2-$\sigma$
bounds as in Figs.~\ref{fig-100kmN}b.  { (a) Non-turbulent case-A
simulations, where initially all the mass is in $D=2$~m particles and
the 100~km objects are introduced progressively over 2~My. Here, the
velocity dispersion of the 2~m-particles evolves according to the
damping (collisional \& gas drag) and viscous stirring equations.  (b)
The same as (a) but with different assumptions on the dynamical
evolution of the particles. The dashed curve refers to the case where
the radial drift speed of the 2~m particles due to gas drag is also
taken into account. The solid curve refers to the case where the
particles are assumed to be tracers of much smaller bodies, fully
coupled with the gas.} (c) Non-turbulent case-B simulation, where
equal masses of 2~m-particles and 100km-planetesimals are introduced
progressively over 2~My, up to a total of 3.2~$M_\oplus$. (d) Case-A
simulations where the velocity dispersion of particles and
planetesimals is stirred by turbulent scattering. The solid curve is
for $\gamma=2\times 10^{-4}$, {the dashed and dotted curves for
$\gamma=2\times 10^{-5}$; in the case shown by the dotted curve we
  impose that the minimal eccentricity of the $D=2$~m particles cannot
decrease below 0.001 and the inclination below $5\times 10^{-4}$.}}
\label{fig-100km} 
\end{figure}

The solid black line in Fig.~\ref{fig-100km}a shows the final SFD
obtained in case-A. The availability of small bodies promotes runaway
growth among the 100-km planetesimals introduced at early times into
the simulation.  This leads to {  very distinctive signatures in the
resulting SFD: a steep fall-off above the input size of the
planetesimals; the presence of very large planetary embryos; a very
shallow slope at moderate sizes (in this case, from slightly more than
100 to several 1,000km) and an overall deficit of objects in this
size-range. As a result, the SFD that does not match that of the
reconstructed main belt even within the 2-$\sigma$ boundaries}.

For completeness, we present {  in Fig.~\ref{fig-100km}b} two
additional variants of this nominal simulation.  In one, inspired by
the Cuzzi et al. work, we assume that our 2~m-particles are tracers
for chondrule-size objects. Chondrules would be strongly coupled with
the gas, so we assume, for simplicity sake, that the particles are
{\it perfectly} coupled with the gas. In practice, instead of letting
our particles evolve in velocity space according to the
damping/stirring equations of {\tt Boulder} (as in the nominal
simulation), we force them to have the same velocity of the gas
(i.e. 60~m/s) relative to Keplerian orbits. The result is illustrated
by the black solid curve. In the second variant, inspired by the
Johansen et al. work, we assume that our 2~m-particles are tracers for
meter-size boulders. These objects should migrate very quickly towards
the Sun by gas drag. We neglect radial migration in {\tt Boulder}
because the annulus that we consider (2--3 AU) is too narrow. This is
equivalent to assuming that the bodies that leave the annulus through
its inner boundary are substituted by new bodies drifting into the
annulus through its outer boundary. The drift speed, however, should
be included in our calculation of the relative velocities of particles
and planetesimals.  Accordingly, we add a 100m/s radial component to
the velocities of all our tracers. The result is illustrated by the
black dashed curve. We find the dashed and solid curves are very
similar to the solid curve of panel~a. Thus, none of the considered
effects appear to have much effect in changing the final SFD. Based on
this, we believe it will be reasonable to neglect these corrections to
the velocity of our particles in the remaining simulations.  This
reduces the number of cases to be investigated and simplifies our
discussion.

The solid black line in Fig.~\ref{fig-100km}c shows the final SFD
obtained in case-B. Again, the signature of runaway growth, triggered
by the availability of a large amount of mass in small particles, is
highly visible. Consequently, the SFD does not match at all that of
the reconstructed main belt. In particular, it shows a strong deficit
of 100-1,000~km bodies.

In order to get a better match with the SFD of the main belt, we would
need to suppress/reduce the signature of runaway growth. One potential
way to do this is to enhance the dispersion velocities of small bodies
via turbulent scattering (see sect.~\ref{turbuN}).  Thus, we proceed
to the inclusion of this effect, which addresses issue (II) mentioned
above in this section.

{ We start by assuming that the parameter $\gamma=2\times 10^{-4}$
  (relatively small compared to expectations).  For the 2~m-particles
  we assume initial eccentricities and inclinations that are the
  equilibrium values obtained by balancing the stirring effect of the
  turbulent disk with the damping effects due to gas drag and mutual
  collisions ($e\sim 2i\sim 7\times 10^{-5}$).  For the
  100km-planetesimals that are injected in the simulation, we assume
  that eccentricity and inclination are 1/2 of their equilibrium
  values (accounting also for tidal damping; Ida et al., 2008). This
  means $e=0.005$ and $i=e/2$. A simulation of the evolution of the
  eccentricity/inclination of a 100~km-planetesimal in a turbulent
  disk shows that these values are achieved in $\sim 100,000$~y
  starting from a circular orbit within the disk's mid-plane.} The
  solid black curve in Fig.~\ref{fig-100km}d shows the result for the
  case-A simulation { with this settings}.  Even with this small
  amount of turbulent scattering, the accretion is strongly inhibited
  and the largest objects do not exceed $D=200$~km. { In a second
  simulation, we decreased $\gamma$ by a factor of 10, {as well as
  the initial eccentricities and inclinations.} This } makes turbulent
  scattering so weak that runaway growth turns back on, making the
  final SFD similar to those shown in panel~b. {We remark, though,
  that the initial eccentricity and inclination of the particles are
  much smaller than what one might expect, due to simple diffusion due
  to local turbulence (Cuzzi and Weidenschilling, 2006). This might
  have favored runaway growth. In reality, turbulent
  diffusion should prevent the eccentricities and inclinations of
  small  bodies to become smaller than $e\sim 2i\sim 10^{-3}$ (Cuzzi,
  private communication). Thus, we did a third simulation, still
  adopting $\gamma=2\times 10^{-5}$, but imposing that $e$ and $i$ of
  our particles/tracers do never decrease below these minimal
  values. The result is shown by the dotted curve. Runaway growth is
  now less extreme than in the previous case (the slope of the SFD
  just above $D=100$~km is shallower and the final embryos are smaller), but
  it is still effective.  Again, the final SFD is inconsistent with the
  asteroid belt constraints. Thus, we conclude that} the
  accretion process in the presence of a large mass of small particles
  is very {  sensitive to the effects of} turbulent scattering: if
  turbulent fluctuations are too violent, accretion is shut off; if
  they are too weak, runaway growth occurs. In both cases, no match
  can be found for the reconstructed main belt.

We conclude from these simulations that the initial planetesimal SFD
had to span a significant range of sizes; our best guess would be
upwards from 100~km. In the next two sections we will try to constrain
the size ($D_{\rm max}$) of the largest initial planetesimals and the
SFD in the 100~km--$D_{\rm max}$ range that are necessary to achieve a
final SFD consistent with that of the reconstructed belt.

\section{Accretion from 100--500~km planetesimals}  

Here we start with planetesimals in the $D = 100$-500~km diameter
range, with an initial SFD whose slope is the same as the one observed
in the reconstructed (and current) SFD of the asteroid belt.

\begin{figure}[t!] 
\centerline{
\includegraphics[height=5.cm]{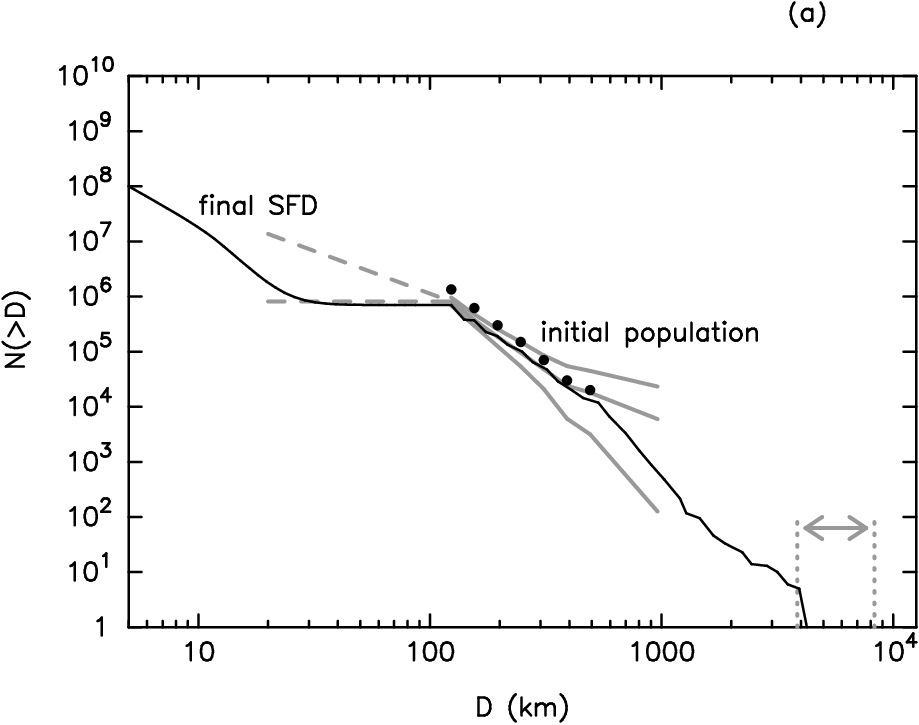} 
\includegraphics[height=5.cm]{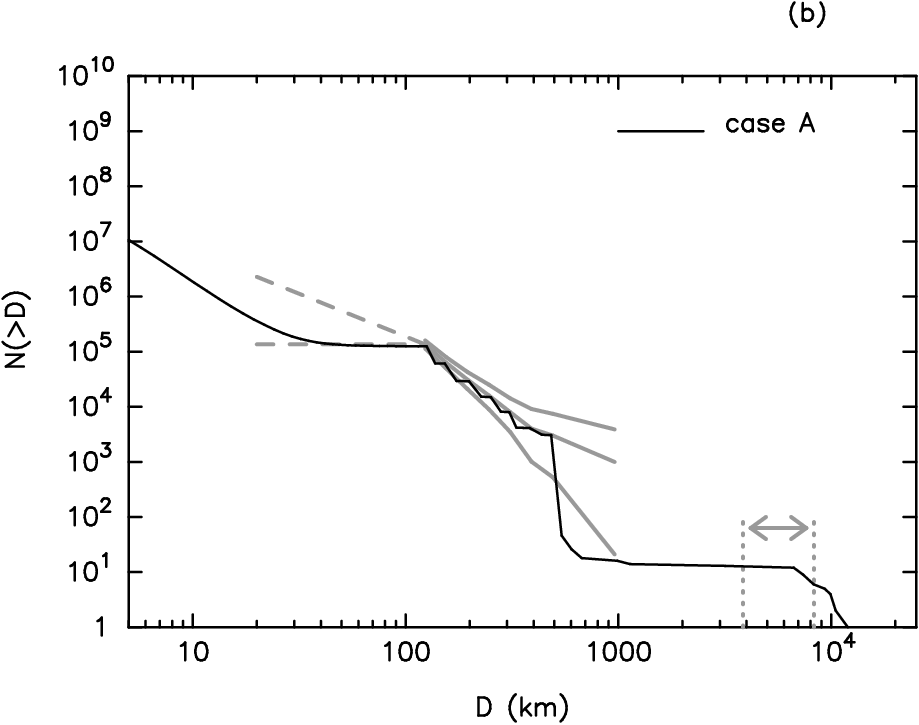}
} 
\centerline{\includegraphics[height=5.cm]{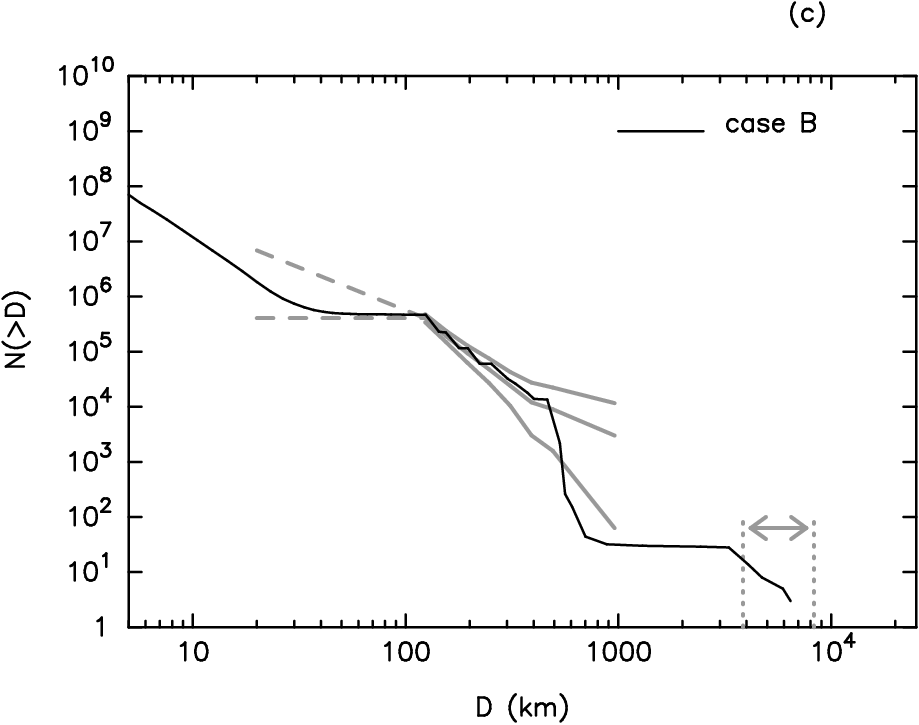} 
\includegraphics[height=5.cm]{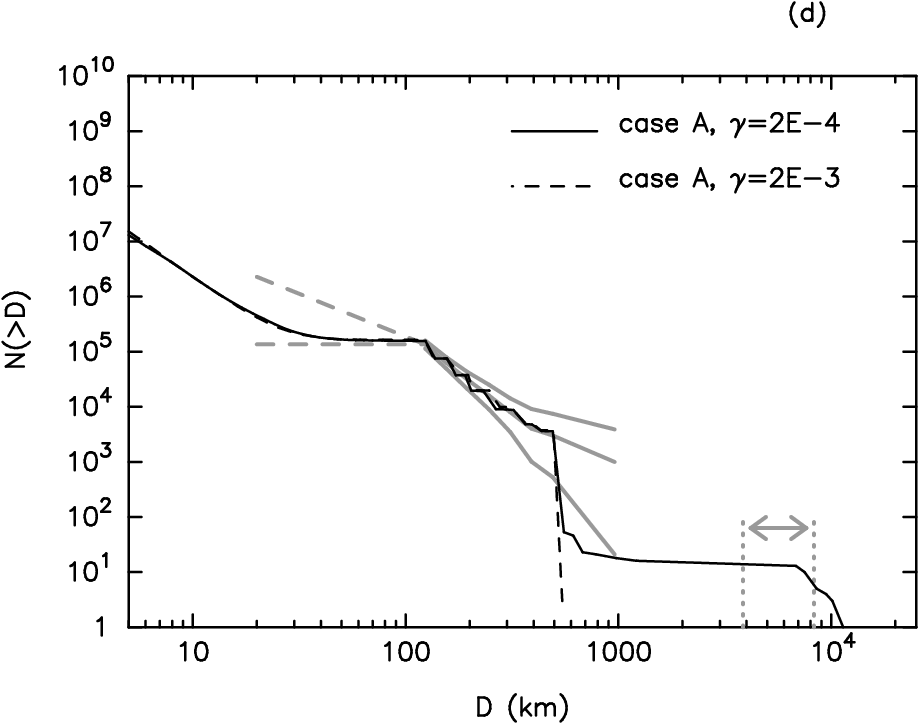}
} 
\vspace*{-.3cm} 
\caption{\small Like Fig.~\ref{fig-100km}, but starting from initial
planetesimals in the 100-500~km range and a main belt-like SFD in this
size interval, as shown by the filled dots in panel (a). {  The
three solid gray curves reproduce the reconstructed SFD and its
2-$\sigma$ bounds (see Fig.~\ref{SFDorig}), rescaled so that the total
number of $D>100$km objects matches that obtained in each
simulation.}} 
\label{fig-100-500} 
\end{figure}

In the first simulation, all the planetesimals are input at $t=0$, as
in the simulation of Fig.~\ref{fig-100kmN}a. In order to place
1.6~$M_\oplus$ in these bodies, we have to assume they were $\sim
10,000$~times more numerous than current asteroids in the same size
range. {  As in the previous sections, no turbulent scattering is
taken into account in this first simulation}. The final SFD is shown
in Fig.~\ref{fig-100-500}a. An important result is that the slope of
the input SFD is preserved to the end of the simulation. The turn-over
of the final SFD at $D\sim 100$~km is recovered and a few Lunar-mass
embryos are produced. {  Notice, though, that the final SFD shows a
sharp break at the initial planetesimals' maximum size ($D>500$~km);
for sizes larger than this threshold, the slope is steeper than the
initial slope in the 100-500km range. The observed SFD of the asteroid
(middle grey solid line in the figure) does not show this behavior. As
discussed in section~2, however, the observed SFD is determined by a
single object (i.e. Ceres) and therefore the determination of the SFD
of the reconstructed belt is affected by small number statistics. With
a 95.8\% probability, the post-accretion SFD of the asteroid belt
should be confined between the upper and lower solid gray curves shown
in the figure.  We find the final SFD in our simulation does fulfill
this requirement very well.}

{  One might be tempted to claim success on the basis of this
simulation, but we caution that this run is overly simplistic for the
reasons that we enumerated in the previous section: we
assumed that (I) all planetesimals are introduced at time = 0 My and
(II) turbulent scattering was not taken into account. We lift these
approximations below.}

A simulation conducted with the case-A set-up discussed in the
previous section (Fig.~\ref{fig-100-500}b) {  exacerbates the break
of the SFD at $D\sim 500$km}. This is because the large planetesimals
introduced early in the simulation efficiently gobble up the small
bodies and form embryos more massive than Mars via runaway growth. The
final SFD is in fact typical of this growth mode (see sect.~4): it shows
a steep slope just above the initial size of the planetesimals
and a deficit of $\sim$~1,000~km objects. {  The final SFD goes
outside of the 2-$\sigma$ boundaries of the reconstructed belt's SFD,
with only 13 bodies with diameters between 500 and 2,000~km (5 bodies
if $D$ is restricted to be larger than 530~km).  Given the dynamical
depletion factor of 1,000 required to bring the number of $D>100$km
objects to the current number, the probability that Ceres survived is
only 1.3\% or 0.5\%. Thus, even accounting for small number statistics
in the observed asteroid SFD, this simulation is highly unlikely to
reproduce the reconstructed main belt.}

The result of a simulation conducted with the case-B set-up discussed
in the previous section is shown in Fig.~\ref{fig-100-500}c. Again, we
see in the final SFD the distinctive signature of runaway growth{ 
with a sharp break in the slope of the SFD at $D\sim 500$km. Hence,
the considerations for the previous run also apply in this case}.

The runs accounting for turbulent scattering are shown in
Fig.~\ref{fig-100-500}d. As in the previous section, all simulations
are conducted within the framework of the case-A set-up. The solid
curve refers to the simulation where $\gamma=2\times 10^{-4}$. Unlike
the run in Fig.~\ref{fig-100km}d, this value of turbulence strength
does not inhibit accretion in this case, such that the signature of
runaway growth is evident in the final SFD {  (this simulation does
not produce a good match to the main belt SFD, as in the cases of
panels~b and c)}. We defer to section \ref{turbu} a discussion on
which values of $\gamma$ allow accretion as a function of planetesimal
sizes. Conversely, if $\gamma$ is increased to $\gamma=2\times
10^{-3}$, accretion is inhibited and the final SFD above the
$D=500$~km drops vertically.  As in the previous section, we conclude
that the accretion process is very unstable with respect to turbulent
scattering: if turbulence is too violent, accretion is shut off; if it
is too weak, runaway growth occurs.

Thus, from all our runs, we conclude that {  it is unlikely that the
asteroid belt SFD can be reproduced } if we start with planetesimals
solely in the 100--500~km size range.  Our insights from these runs
also suggest that {  reducing the size of the largest initial
planetesimals is only going to make the match more problematic.} Thus,
we argue that the initial planetesimals had to span the full
100--1,000km range, with a power law slope similar to that of the main
belt SFD. In the next section we check whether this initial
planetesimal distribution does indeed lead to a final distribution
matching all asteroid belt constraints.

\section{Accretion from 100--1,000~km planetesimals}

In this section, we redo all the runs presented in the previous section
but extend the size distribution of the initial planetesimals up to
Ceres-size bodies ($D \sim 1000$~km).

\begin{figure}[t!] 
\centerline{
\includegraphics[height=5.cm]{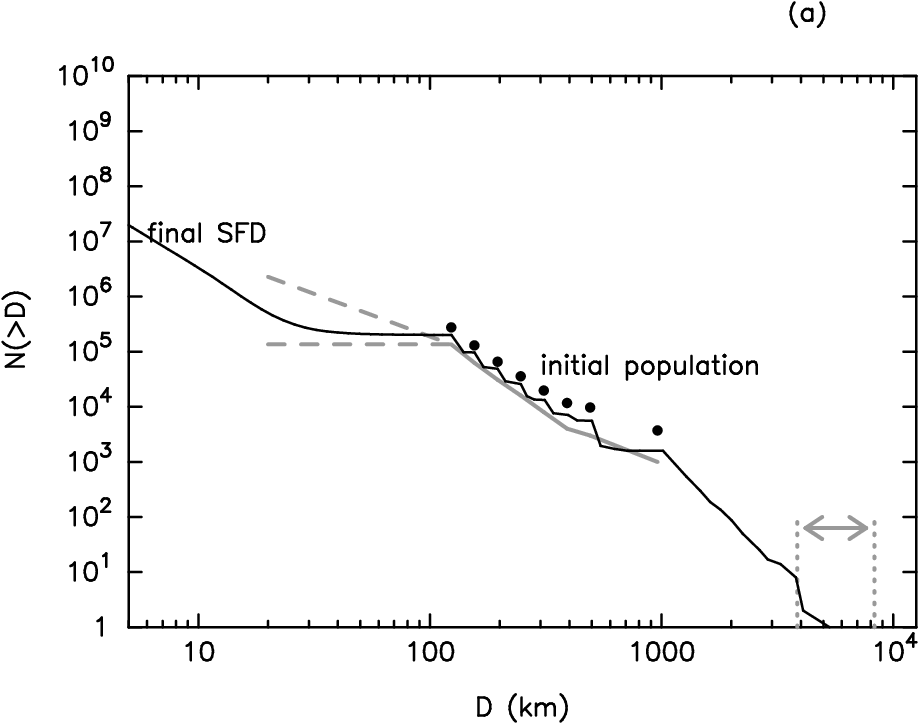} 
\includegraphics[height=5.cm]{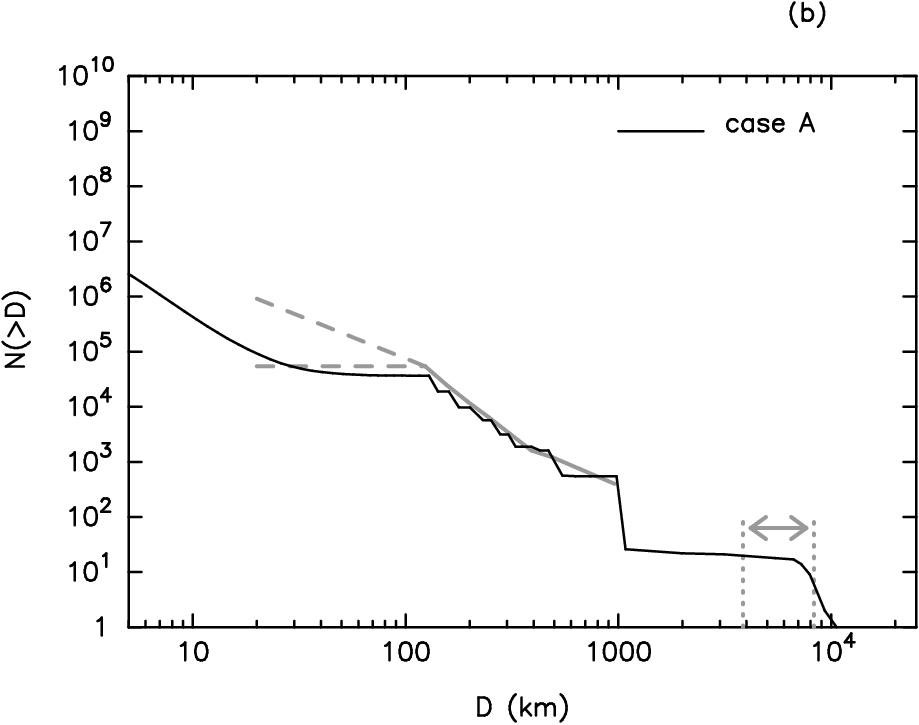}
} 
\centerline{
\includegraphics[height=5.cm]{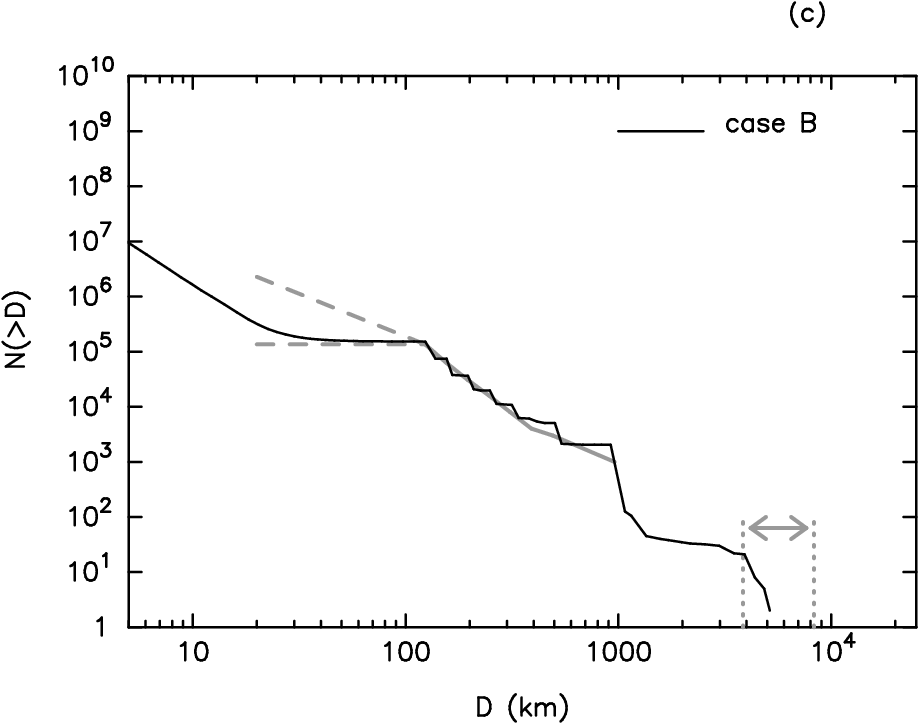} 
\includegraphics[height=5.cm]{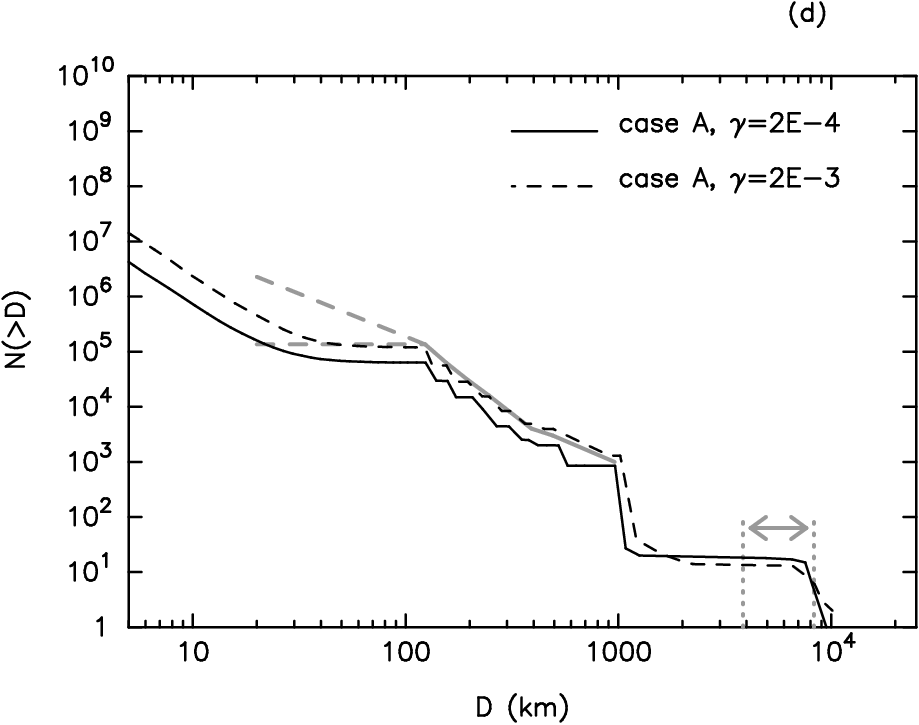}
} 
\vspace*{-.3cm} 
\caption{\small Like Fig.~\ref{fig-100-500}, but
starting from initial planetesimals in the 100-1,000~km range and an
SFD slightly shallower than that of the main belt in this size
interval, as shown by the bullets in panel (a). The solid gray line
reproduces the {  current SFD of the main belt, rescaled so
that the total number of $D>100$km objects matches that obtained in
each simulation. (400 in panel (b), 1,000 in all other panels).}}
\label{fig-1000} 
\end{figure}

In our nominal simulation, {  which does not include turbulent
scattering,} we start from a population of initial 100-1,000~km
planetesimals in quantities that are 2,000-4,000 times the main belt
population from the small to the large end (i.e., they have a SFD
slightly shallower than the current main belt SFD). Because the total
mass of this population is only 0.9~$M_\oplus$, instead of the
1.6~$M_\oplus$ used in all other previous simulations, we place the
remaining mass (0.7~$M_\oplus$) in $D = 2$~m bodies. These bodies are
treated like normal planetesimals in this run: they can accrete or
break in mutual collisions. We find that $\sim 10$\% of the meter-size
bodies coagulate with the large planetesimals, while the rest are
eliminated by collisional grinding. The final SFD, shown by the black
solid curve in Fig.~\ref{fig-1000}a, is now consistent with properties
(i)-(iv) of the reconstructed post-accretion asteroid belt.

Our results, once properly scaled, are consistent with those found by
direct $N$-body simulations also starting with large planetesimals
(e.g. Kokubo and Ida, 2000; KI00). The simulation in KI00 lasts
500,000~y in an annulus centered on 1~AU.  This is equivalent to our
simulations where we examine what happens over 2~My to an annulus
centered around 2.5~AU.  After 2~My of coagulation, our biggest object
has a mass of $1.5\times 10^{26}$~g. In KI00, its mass is $2\times
10^{26}$~g.  In KI00 there are 7 bodies more massive than $10^{25}$~g
in their $\pm 0.04$~AU wide annulus. In an annulus that is 7.5 times
larger we have 51 bodies more massive than this threshold. More
importantly, KI00 also finds that the SFD of their initial
planetesimal remains essentially unchanged during the simulation (see
their Fig.~8).

Using the case-A and B set-ups (Figs.~\ref{fig-1000}b
and~\ref{fig-1000}c), the final SFDs show the distinct signatures of
runaway growth, though it is more pronounced in case-A than in case-B.
The final planetary embryos are also more numerous and massive than in
the nominal simulation of panel (a). Notice that, in the case-A
simulation, the final number of objects in the input size-range is
smaller than in the other cases.  {  The reason is that every time a
large planetesimal is introduced in the simulation, a number of small
bodies of equivalent total mass is removed. If at some point small
bodies are no longer available because they have been accreted by the
growing embryos, the introduction of new planetesimals is terminated.}

Fig.~\ref{fig-1000}d shows results of simulations accounting for the
excitation of the relative velocities due to turbulent scattering. We
assume $\gamma=2\times 10^{-3}$, which was inhibiting accretion in the
cases with $D_{\rm max}\le 500$km (see Fig.~\ref{fig-100-500}d). Here
($D_{\rm max}=$1,000km) we find that this level of turbulence strength
has little effect on the accretion process. The reasons for this are
discussed in sect.\ref{turbu}.

In summary, all the simulations {  shown in Fig.~\ref{fig-1000}}
give results that are consistent with the reconstructed SFD of the
main belt. The apparent robustness of our results, as opposed to the
systematic failures {  or improbable matches} obtained in the
previous sections, gives us increased confidence that the initial
planetesimal SFD had to span sizes ranging from 100~km up to (at
least) Ceres-size objects.  We argue the initial slope of the initial
planetesimals also had to be similar to the one currently observed in
the main belt population.

\subsection{Note on collisional coagulation and external velocity
  excitation} 
\label{turbu}  

We have shown in Figs.~ \ref{additional}f and \ref{fig-100km}d that if
the input planetesimals are not larger than 100~km, the introduction
of turbulent scattering with $\gamma=2\times 10^{-4}$ causes an
effective negation of the accretion process. However, if the input
planetesimals have sizes ranging from 100 to 500~km
(Fig.~\ref{fig-100-500}d) or 1,000~km (Fig.~\ref{fig-1000}d), the same
turbulent strength does not change the outcome of the simulation with
respect to the case where no turbulent scattering is
included. Similarly, $\gamma=2\times 10^{-3}$ aborts accretion if the
initial planetesimals are not larger than 500~km
(Fig.~\ref{fig-100-500}d) but not in the case where they are
Ceres-size (Fig.~\ref{fig-1000}d).

Turbulent scattering provides an `external' excitation of the velocity
dispersion of the planetesimals. By `external', we do not mean
generated by the interaction among the planetesimals themselves. In
the simulations, and in our discussion below, what matters is the
magnitude of this external excitation and not the process that causes
it. Thus, other forms of excitation, such as, for example,
gravitational stirring from Jupiter's forming core can be considered
as well.

In collisional coagulation, the key factor is the ratio between the
escape velocity from the largest planetesimals and the dispersion
velocity of the bodies carrying the bulk of the total mass, relative to
those planetesimals. In absence of external excitation mechanisms,
the former is always larger than, or of the same order of, the
latter. The first case leads to runaway growth; the second to
oligarchic growth. If an external excitation is present, the velocity
dispersion can become much larger than the escape velocities. This
slows down the coagulation process considerably and effectively ends
it.

For instance, for $\gamma=2\times 10^{-4}$, turbulent excitation
pushes 100~km bodies to eccentricities of $e\sim 0.01$. At 2.5~AU,
this corresponds to a velocity of $\sim 200$~m/s relative to a local
circular orbit. This value is larger than the escape velocity from a
100~km object ($\sim 50$m/s, assuming a bulk density of 2g/cm$^3$) but
is smaller than the escape velocity from a 500~km or 1,000~km object
($\sim 250$--500~m/s). This explains why accretion is aborted in the
first case but not in the other cases, as shown in the simulations
presented above.

The velocity excitation scales linearly with $\gamma$ in first
approximation. Thus, by the argument described above, one would
predict that, in the case with initial planetesimals up to 1,000~km in
size, collisional coagulation is severely inhibited if $\gamma>
7\times 10^{-4}$ because this value would give a velocity dispersion
on the order of $\sim$500~m/s, comparable to the escape velocity from
a 1,000~km object. In reality, we have seen in Fig.~\ref{fig-1000}d
that collisional coagulation is still effective in case-A even for
$\gamma=2\times 10^{-3}$. This is due to the fact that, in case-A, all
the mass is initially in small bodies whose velocity excitation is
reduced due to gas drag and mutual collisions (Ida et al., 2008); in
turn these small bodies damp the velocity dispersion of the large
planetesimals by dynamical friction. We have checked that, if
turbulent excitation is introduced into a simulation with our nominal
set-up (like that of Fig.~\ref{fig-1000}a), collisional coagulation is
indeed negated for $\gamma> 7\times 10^{-4}$.

\section{Conclusions}  

The first and most basic step of the accretion of planets is the
creation of planetesimals. Unfortunately, planetesimal formation is
still poorly understood. In the traditional view, planetesimals grow
progressively from coagulations of dust and pebbles to kilometer-sized
objects. Consequently, the simulations of the second step of the
accretion process, that in which collisional coagulation among the
planetesimals leads to the formation of planetary embryos and giant
planet cores, usually starts from a population of kilometer-sized
planetesimals (e.g. Weidenschilling et al., 1997; Kenyon and Bromley,
2006).

However, recent paradigm-breaking work (Johansen et al., 2007; Cuzzi
et al., 2008) showed that planetesimals might form big (100km or
larger) thanks to the self-gravity of small bodies highly concentrated
in the turbulent structures of the proto-planetary disk. If this is
true, then there are no km-sized initial planetesimals and the second
step of accretion has to be somehow affected by this change in
`initial conditions'.

{  In this work, we have assumed the SFD of the `initial planetesimals'
is unknown and we have attempted to constrain it by matching the final
SFD produced by the second step of the accretion process with the SFD
of a reconstructed asteroid belt. More specifically, the `target SFD'
that we try to reproduce with collisional coagulation simulations is
the one that the asteroid belt should have had just prior to it being
dynamical excited and depleted of material. The large body of work on
the past history of the asteroid belt, which was reviewed in section
2, allows us to define the shape and size of this target SFD.

While it is impossible to prove a negative result, we believe we have
run enough simulations to understand the response of the
coagulation process to various initial and environmental conditions.}
Based on these results, we find it likely that the SFD of the asteroid
belt cannot be reproduced from an initial population of km-sized
planetesimals. It also cannot be reproduced by assuming that the
initial planetesimals had sizes up to some value $D_{\rm max}<${ 
500--}1,000~km. Instead, we find that the reproduction of the asteroid
belt constraints requires that the initial planetesimals had to span
{  the size range from $\sim 100$ to several 100 km, probably up to
1,000km}, and that their initial SFD had a slope {  similar} to that
of the current SFD of asteroids in the same size-range. Curiously,
this result is reminiscent of the original intuition by Kuiper (1958)
that the original asteroid size distribution had to have a Gaussian
shape centered around 100~km.

Our result provides support for the idea that planetesimals formed big
(Johansen et al., 2007; Cuzzi et al., 2008). The precise 
process that formed these big planetesimals is still an open issue. Our
findings (size range and SFD slope of the initial planetesimals)
should help constrain the planetesimal formation models.

We have also shown that, if the initial planetesimals can be as big as
1,000~km, the subsequent collisional coagulation process leading to
the formation of planetary embryos is not seriously affected by the
excitation of eccentricities and inclinations due to the turbulence in
the disk. This may provide a possible solution for the problem of
planet formation in turbulent disks (Nelson, 2005; Ida et al., 2008).

Our results also help us explain several interesting mysteries about
small body evolution across the solar system. For example, if we
assume the asteroid belt was initially deficient in $D < 100$~km
asteroids, its early collisional activity may have been much lower
than previously thought. {Thus, the constraint provided by the
uniqueness of Vesta's large basin (i.e. that the asteroid belt hosted
cumulatively over its history a population of $D>35$km objects
equivalent to 1,000 bodies for $4$~Gy; see sect.~\ref{collNO}) could
be fulfilled even if the number of ``big'' asteroids (e.g. $D>100$km)
remained larger than now for some time (for instance up to the
LHB). The initial deficit of small asteroids could} also explain the
paucity of meteorite shock degassing ages recorded between 4.1-4.4~Gy
ago (Kring and Swindle, 2008) and, for extra-solar systems, the
deficit of hot dust observed in young proto-planetary disks
(Silverstone et al., 2006). Moreover, if planetesimals formed in the
same way in the Kuiper belt, it is likely that the turn-over observed
in its SFD at $D\sim 100$~km (Bernstein et al., 2004) is also a
signature of accretion and not one of collisional grinding, unlike
what it is usually assumed (e.g. Kenyon and Bromley, 2004; Pan and
Sari, 2005).

Finally, we have shown that the sudden appearance of large
planetesimals in a massive disk of small bodies boosts runaway
accretion of large objects (see the case-A/B simulations in sections 4
and 5). This result might help in solving the problem of the formation
of the Jovian planet cores, one of the major open issues in planetary
science.

\vskip 20pt {\centerline {\bf Acknowledgments}}  

This work was done while the first author was on sabbatical at
SWRI. A.M. is therefore grateful to SWRI and CNRS for providing the
opportunity of this long term visit and for their financial support. {
This paper had three reviewers: J. Chambers, J. Cuzzi and
S. Weidenschilling. The challenges that they set made this manuscript
evolve over more than one year from a Nature-size letter (although
with a long SI) to a thesis-size monograph (blame them if you thought
that this paper is too long!). However, these challenges also made
this work more extensive, robust and -hopefully- convincing, and we
thank the reviewers for this.} We also thank Scott Kenyon for a
friendly review of an early version of the electronic supplement of
this paper, describing our code and its tests.

\vskip 20pt {\centerline {\bf References}}

\begin{itemize}

\item[$\bullet$] Agnor, C.~B., Canup, 
R.~M., Levison, H.~F.\ 1999. On the Character and Consequences of Large 
Impacts in the Late Stage of Terrestrial Planet Formation.\ {\it Icarus} 142, 
219-237.

\item[$\bullet$] Alexander, 
C.~M.~O.~'., Grossman, J.~N., Ebel, D.~S., Ciesla, F.~J.\ 2008.\ The 
Formation Conditions of Chondrules and Chondrites.\ Science 320,
1617. 

\item[$\bullet$] Alibert, Y., Mordasini, C., Benz, W.\ 2004.\ Migration and giant
planet formation.\ Astronomy and Astrophysics 417, L25-L28.

\item[$\bullet$] Alibert, Y., Mousis, 
O., Mordasini, C., Benz, W.\ 2005.\ New Jupiter and Saturn Formation Models 
Meet Observations.\ Astrophysical Journal 626, L57-L60. 

\item[$\bullet$] Baruteau, C. 2009. Protoplanetary migration in
  turbulent isothermal disks. Astrophysical Journal, submitted. 

\item[$\bullet$] Benz, W., Asphaug, 
E.\ 1999. Catastrophic Disruptions Revisited.\ {\it Icarus} 142, 5-20.

\item[$\bullet$] Bernstein, G.~M., 
Trilling, D.~E., Allen, R.~L., Brown, M.~E., Holman, M., Malhotra, R.\ 
2004.\ The Size Distribution of Trans-Neptunian Bodies.\ Astronomical 
Journal 128, 1364-1390. 

\item[$\bullet$] Blum, J., and 26 
colleagues 2000.\ Growth and Form of Planetary Seedlings: Results from a 
Microgravity Aggregation Experiment.\ Physical Review Letters 85, 
2426-2429. 

\item[$\bullet$] Bogard, D.\ 1995. Impact ages 
of meteorites: A synthesis.\ {\it Meteoritics} 30, 244. 

\item[$\bullet$] Bottke, W.~F., Nolan, 
M.~C., Greenberg, R., Kolvoord, R.~A.\ 1994.\ Velocity distributions among 
colliding asteroids.\ Icarus 107, 255-268. 

\item[$\bullet$] Bottke, W.~F. Durda, 
D.~D., Nesvorn{\'y}, D., Jedicke, R., Morbidelli, A., Vokrouhlick{\'y}, D., 
Levison, H.\ 2005.\ The fossilized
size distribution of the main asteroid belt.\ {\it Icarus} 175,
111-140.

\item[$\bullet$] Bottke, W.~F., Durda, 
D.~D., Nesvorn{\'y}, D., Jedicke, R., Morbidelli, A., Vokrouhlick{\'y}, D., 
Levison, H.~F.\ 2005b.\ Linking the collisional history of the main asteroid 
belt to its dynamical excitation and depletion.\ Icarus 179, 63-94. 

\item[$\bullet$] Britt, D.~T., Yeomans, 
D., Housen, K., Consolmagno, G.\ 2002. Asteroid Density,
Porosity, and Structure.\ {\it Asteroids III} 485-500.

\item[$\bullet$] Chambers, 
J.~E., Wetherill, G.~W., 1998.\ Making the Terrestrial Planets: N-Body 
Integrations of Planetary Embryos in Three Dimensions.\ {\it Icarus} 136, 
304-327.

\item[$\bullet$] Chambers, J.~E.\ 2001.
Making More Terrestrial Planets.\ {\it Icarus} 152, 205-224. 

\item[$\bullet$] Chambers, J.~E., Wetherill, G.~W.\  2001. Planets
  in the asteroid belt.\ {\it Meteoritics and Planetary Science} 36,
  381-399.

\item[$\bullet$] Chambers, J.\ A 
semi-analytic model for oligarchic growth.\ {\it Icarus} 180, 496-513
(2006).

\item[$\bullet$] Ciesla, F.~J., Hood, 
L.~L.\ 2002. The Nebular Shock Wave Model for Chondrule Formation: Shock 
Processing in a Particle-Gas Suspension.\ {\it Icarus} 158, 281-293. 

\item[$\bullet$] Colwell, J.~E., 
Taylor, M.\ 1999.\ Low-Velocity Microgravity Impact Experiments into 
Simulated Regolith.\ Icarus 138, 241-248. 

\item[$\bullet$] Connolly, H.~C., 
Jr., Love, S.~G.\ 1998. The formation of chondrules: petrologic tests of 
the shock wave model..\ {\it Science} 280, 62-67. 

\item[$\bullet$] {Cuzzi, 
J.~N., Weidenschilling, S.~J.\ 2006.\ Particle-Gas Dynamics and Primary 
Accretion.\ Meteorites and the Early Solar System II 353-381.}

\item[$\bullet$] Cuzzi, J.~N., Hogan, 
R.~C., Paque, J.~M., Dobrovolskis, A.~R.\ 2001.\ Size-selective 
Concentration of Chondrules and Other Small Particles in Protoplanetary 
Nebula Turbulence.\ Astrophysical Journal 546, 496-508. 

\item[$\bullet$] Cuzzi, J.~N., Hogan, 
R.~C., Shariff, K.\ 2008.\ Toward Planetesimals: Dense Chondrule Clumps in 
the Protoplanetary Nebula.\ Astrophysical Journal 687, 1432-1447. 

\item[$\bullet$] Davis, D.~R.,  Chapman, 
C.~R., Weidenschilling, S.~J., Greenberg, R.\ 1985.  Collisional history of
asteroids: Evidence from Vesta and the Hirayama families.\ {\it Icarus} 63,
30-53.

\item[$\bullet$] Davis, D.~R., Durda, 
D.~D., Marzari, F., Campo Bagatin, A., Gil-Hutton, R.\ 2002. Collisional 
Evolution of Small-Body Populations.\ {\it Asteroids III} 545-558. 

\item[$\bullet$] Desch, S.~J., Connolly, H.~C., Jr.\ 2002. A model of the
thermal processing of particles in solar nebula shocks: Application to
the cooling rates of chondrules.\ {\it Meteoritics and Planetary Science}
37, 183-207.

\item[$\bullet$] Dominik, C., 
Tielens, A.~G.~G.~M.\ 1997.\ The Physics of Dust Coagulation and the 
Structure of Dust Aggregates in Space.\ Astrophysical Journal 480, 647. 

\item[$\bullet$] Dominik, C., Blum, J., 
Cuzzi, J.~N., Wurm, G.\ 2007.\ Growth of Dust as the Initial Step Toward 
Planet Formation.\ Protostars and Planets V 783-800. 

\item[$\bullet$] Durda, D.~D., Greenberg, 
R., Jedicke, R.\ 1998.\ Collisional Models and Scaling Laws: A New 
Interpretation of the Shape of the Main-Belt Asteroid Size Distribution.\ 
{\it Icarus} 135, 431-440. 

\item[$\bullet$] Durda, D.~D., Bottke, 
W.~F., Enke, B.~L., Merline, W.~J., Asphaug, E., Richardson, D.~C., 
Leinhardt, Z.~M.\  \ 2004. The formation
of asteroid satellites in large impacts: results from numerical
simulations.\ {\it Icarus} 170, 243-257.

\item[$\bullet$] Durda, D.~D., Bottke, 
W.~F., Nesvorn{\'y}, D., Enke, B.~L., Merline, W.~J., Asphaug, E., 
Richardson, D.~C.\ 2007.\ Size frequency distributions of fragments from 
SPH/N-body simulations of asteroid impacts: Comparison with observed 
asteroid families.\ Icarus 186, 498-516. 

\item[$\bullet$] Goldreich, P., 
Lithwick, Y., Sari, R.\ 2004.\ Final Stages of Planet Formation.\ 
Astrophysical Journal 614, 497-507. 

\item[$\bullet$] Gomes, R., Levison, 
H.~F., Tsiganis, K., Morbidelli, A.\ 2005.\ Origin of the cataclysmic Late 
Heavy Bombardment period of the terrestrial planets.\ {\it Nature} 435,
466-469. 

\item[$\bullet$] Greenberg, R., 
Hartmann, W.~K., Chapman, C.~R., Wacker, J.~F.\ 1978.\ Planetesimals to
planets - Numerical simulation of collisional evolution.\ {\it Icarus}
35, 1-26.

\item[$\bullet$] Greenzweig, 
Y., Lissauer, J.~J. 1992. Accretion rates of protoplanets. II - Gaussian 
distributions of planetesimal velocities.\ {\it Icarus} 100, 440-463.

\item[$\bullet$] Grieve, R. A. and Shoemaker, E. M. 1994. The record of
past impacts on  Earth. In  {\it Hazards Due to Comets and Asteroids} 
417--462.

\item[$\bullet$] Haisch K.E., Lada E.A. and Lada
 C.J. 2001.  Disk frequencies and lifetimes in young
 clusters. {\it Astroph. J.}, 553, L153--156.

\item[$\bullet$] Hayashi, C.\ 1981. Structure of the Solar
Nebula, Growth and Decay of Magnetic Fields and Effects of Magnetic
and Turbulent Viscosities on the Nebula.\ {\it Progress of Theoretical
Physics Supplement} 70, 35-53.

\item[$\bullet$] Ida, S., Makino, J.\ 1993.
Scattering of planetesimals by a protoplanet - Slowing down of 
runaway growth.\ {\it Icarus} 106, 210-218.

\item[$\bullet$] Ida, S., Lin, D.~N.~C.\ 
2004a.\ Toward a Deterministic Model of Planetary Formation. I. A Desert in 
the Mass and Semimajor Axis Distributions of Extrasolar Planets.\ 
Astrophysical Journal 604, 388-413. 

\item[$\bullet$] Ida, S., Lin, D.~N.~C.\ 
2004b.\ Toward a Deterministic Model of Planetary Formation. II. The 
Formation and Retention of Gas Giant Planets around Stars with a Range of 
Metallicities.\ Astrophysical Journal 616, 567-572. 

\item[$\bullet$] Ida, S., Guillot, T., 
Morbidelli, A. 2008. Accretion and destruction of planetesimals in 
turbulent disks.{\it Astroph. J.} 686, 1292-1301.

\item[$\bullet$] Jedicke, R., Larsen, 
J., Spahr, T.\ 2002. Observational Selection Effects in Asteroid Surveys.\ 
{\it Asteroids III} 71-87.

\item[$\bullet$] Johansen, A., Henning, 
T., Klahr, H.\ 2006.\ Dust Sedimentation and Self-sustained 
Kelvin-Helmholtz Turbulence in Protoplanetary Disk Midplanes.\ 
Astrophysical Journal 643, 1219-1232. 

\item[$\bullet$] Johansen, A., Oishi, 
J.~S., Low, M.-M.~M., Klahr, H., Henning, T., Youdin, A.\ 2007.\ Rapid 
planetesimal formation in turbulent circumstellar disks.\ Nature 448, 
1022-1025. 

\item[$\bullet$] Krasinsky, G.~A., 
Pitjeva, E.~V., Vasilyev, M.~V., Yagudina, E.~I.\ 2002. Hidden Mass in the 
Asteroid Belt.\ {\it Icarus} 158, 98-105.

\item[$\bullet$] Kempf, S., Pfalzner, S., 
Henning, T.~K.\ 1999.\ N-Particle-Simulations of Dust Growth. I. Growth 
Driven by Brownian Motion.\ Icarus 141, 388-398. 

\item[$\bullet$] Kenyon, S.~J., Luu,
J.~X.\ 1999. Accretion in the Early Outer Solar System.\ {\it Astrophysical
Journal} 526, 465-470.

\item[$\bullet$] Kenyon, S.~J.,
Bromley, B.~C.\ 2001. Gravitational Stirring in Planetary Debris Disks.\
{\it Astronomical Journal} 121, 538-551.

\item[$\bullet$] Kenyon, S.~J., 
Bromley, B.~C.\ 2004.\ The Size Distribution of Kuiper Belt Objects.\ 
Astronomical Journal 128, 1916-1926. 

\item[$\bullet$] Kenyon, S.~J., 
Bromley, B.~C.\ 2006. Terrestrial Planet Formation. I. The Transition from 
Oligarchic Growth to Chaotic Growth. {\it Astronomical Journal} 131,
1837-1850.

\item[$\bullet$] Kokubo, E., Ida, S.\ 1998. 
Oligarchic Growth of Protoplanets.\ {\it Icarus} 131, 171-178.

\item[$\bullet$] Kokubo, E., Ida, S.\ 2000.
Formation of Protoplanets from Planetesimals in the Solar Nebula.\ 
{\it Icarus} 143, 15-27.

\item[$\bullet$] Kring, D.~A., 
Swindle, T.~D. 2008. Impact Cratering on the H-Chondrite Parent Body: 
Implications for the Collisional Evolution of the Inner Solar System.\ 
{\it Lunar and Planetary Institute Conference Abstracts} 39, 1305.

\item[$\bullet$] Kuiper, G.~P.\ 1958.\ 
Proceedings of the Celestial Mechanics Conference: Statistics of asteroids 
(abstract).\ Astronomical Journal 63, 412. 

\item[$\bullet$] Laughlin, G.,
Steinacker, A., Adams, F.~C.\ 2004. Type I Planetary Migration with MHD
Turbulence.\ {\it Astrophysical Journal} 608, 489-496.

\item[$\bullet$] {Leinhardt, 
Z.~M., Stewart, S.~T.\ 2009.\ Full numerical simulations of catastrophic 
small body collisions.\ Icarus 199, 542-559.} 

\item[$\bullet$] Mann, I., Grun, E., Wilck, 
M.\ 1996. The Contribution of Asteroid Dust to the Interplanetary Dust 
Cloud: The Impact of ULYSSES Results on the Understanding of Dust 
Production in the Asteroid Belt and of the Formation of the IRAS Dust 
Bands.\ {\it Icarus} 120, 399-407. 

\item[$\bullet$] Merline, W.~J., 
Weidenschilling, S.~J., Durda, D.~D., Margot, J.~L., Pravec, P., Storrs, 
A.~D.\ 2002.\ Asteroids Do Have Satellites.\ Asteroids III 289-312. 

\item[$\bullet$] Minton, D.~A., 
Malhotra, R.\ 2009.\ A record of planet migration in the main asteroid 
belt.\ Nature 457, 1109-1111. 

\item[$\bullet$] Morbidelli, A., Chambers, J., Lunine, J.~I., Petit,
J.~M., Robert, F., Valsecchi, G.~B., Cyr, K.~E.\ 2000.\ Source regions
and timescales for the delivery of water to Earth.\ Meteoritics and
Planetary Science 35, 1309-1320.

\item[$\bullet$]  Nelson, R.~P.\ 2005. On the orbital
evolution of low mass protoplanets in turbulent, magnetised disks.\
{\it Astronomy and Astrophysics} 443, 1067-1085.

\item[$\bullet$] O'Brien, D.~P., 
Greenberg, R.\ 2005. The collisional and dynamical evolution of the 
main-belt and NEA size distributions.\ {\it Icarus} 178, 179-212.

\item[$\bullet$] O'Brien, D.~P., 
Morbidelli, A., Levison, H.~F. 2006. Terrestrial planet formation with 
strong dynamical friction.\ Icarus 184, 39-58.

\item[$\bullet$] O'Brien, D.~P.,
Morbidelli, A., Bottke, W.~F. 2007. The primordial excitation and
clearing of the asteroid belt Revisited.\ {\it Icarus} 191, 434-452.

\item[$\bullet$] Pan, M., Sari, R.\ 2005.\ 
Shaping the Kuiper belt size distribution by shattering large but 
strengthless bodies.\ Icarus 173, 342-348. 

\item[$\bullet$] {Papaloizou, 
J.~C.~B., Nelson, R.~P.\ 2003.\ The interaction of a giant planet with a 
disc with MHD turbulence - I. The initial turbulent disc models.\ Monthly 
Notices of the Royal Astronomical Society 339, 983-992. }

\item[$\bullet$] Petit, J.-M., Morbidelli,
A., Chambers, J.\ 2001.  The Primordial Excitation and Clearing of the
Asteroid Belt.\ {\it Icarus} 153, 338-347.

\item[$\bullet$] Petit, J.-M., Chambers, 
J., Franklin, F., Nagasawa, M.\ 2002. Primordial Excitation and Depletion 
of the Main Belt.\ {\it Asteroids III} 711-723. 

\item[$\bullet$] Podolak, M., Zucker, S.\ 2004.\ A note on the snow line in
protostellar accretion disks.\ Meteoritics and Planetary Science 39,
1859-1868. 

\item[$\bullet$] Pollack, J.~B., 
Hubickyj, O., Bodenheimer, P., Lissauer, J.~J., Podolak, M., Greenzweig, 
Y.\ 1996.\ Formation of the Giant Planets by Concurrent Accretion of Solids 
and Gas.\ Icarus 124, 62-85. 

\item[$\bullet$] Raymond, S.~N., Quinn, 
T., Lunine, J.~I.\ 2004. Making other earths: dynamical simulations of 
terrestrial planet formation and water delivery.\ {\it Icarus} 168, 1-17.

\item[$\bullet$] Raymond, S.~N., Quinn, 
T., Lunine, J.~I.\ 2005. Terrestrial Planet Formation in Disks with 
Varying Surface Density Profiles.\ {\it Astrophysical Journal} 632,
670-676.

\item[$\bullet$] Raymond, S.~N., Quinn, 
T., Lunine, J.~I.\ 2006. High-resolution simulations of the final assembly 
of Earth-like planets I. Terrestrial accretion and dynamics.\ {\it Icarus} 183, 
265-282.

\item[$\bullet$] Raymond, S.~N., Quinn, 
T., Lunine, J.~I. 2007. High-Resolution Simulations of The Final Assembly 
of Earth-Like Planets. 2. Water Delivery And Planetary Habitability.\ 
{\it Astrobiology} 7, 66-84.

\item[$\bullet$] Scott, E.~R.~D.\ 2006.  
Meteoritical and dynamical constraints on the growth mechanisms and 
formation times of asteroids and Jupiter.\ {\it Icarus} 185, 72-82. 

\item[$\bullet$] {Shakura, N.~I., 
Sunyaev, R.~A.\ 1973.\ Black holes in binary systems. Observational 
appearance.\ Astronomy and Astrophysics 24, 337-355.}

\item[$\bullet$] Silverstone, M.~D., 
and 16 colleagues \ 2006. Formation and Evolution of Planetary Systems 
(FEPS): Primordial Warm Dust Evolution from 3 to 30 Myr around Sun-like 
Stars.\ {\it Astrophysical Journal} 639, 1138-1146. 

\item[$\bullet$] {Stewart, S.~T., 
Leinhardt, Z.~M.\ 2009.\ Velocity-Dependent Catastrophic Disruption 
Criteria for Planetesimals.\ Astrophysical Journal 691, L133-L137.} 

\item[$\bullet$] Stone, J.~M., Gammie, 
C.~F., Balbus, S.~A., Hawley, J.~F.\ 2000.\ Transport Processes in 
Protostellar Disks.\ Protostars and Planets IV 589. 

\item[$\bullet$] Strom, R.~G., Malhotra, 
R., Ito, T., Yoshida, F., Kring, D.~A.\ 2005. The Origin of Planetary 
Impactors in the Inner Solar System.\ {\it Science} 309, 1847-1850.

\item[$\bullet$] Thomas, P.~C Binzel, 
R.~P., Gaffey, M.~J., Storrs, A.~D., Wells, E.~N., Zellner, B.~H.\  1997.
Impact excavation on asteroid 4 Vesta: Hubble Space Telescope results.\ 
{\it Science} 277, 1492-1495. 

\item[$\bullet$] Thommes, E.~W., Duncan, 
M.~J., Levison, H.~F.\ 2003. Oligarchic growth of giant planets.\ {\it Icarus} 
161, 431-455.

\item[$\bullet$] Weidenschilling, S.~J.\ 1977. The
 distribution of mass in the planetary system and solar nebula.\
 {\it Astrophysics and Space Science} 51, 153-158.

\item[$\bullet$] Weidenschilling, 
S.~J.\ 1980.\ Dust to planetesimals - Settling and coagulation in the solar 
nebula.\ Icarus 44, 172-189. 

\item[$\bullet$] Weidenschilling, S.~J., Spaute, D., Davis, D.~R., Marzari, F., Ohtsuki, K.\ 1997. Accretional
 Evolution of a Planetesimal Swarm.\ {\it Icarus} 128, 429-455.

\item[$\bullet$] {Weidenschilling, 
S.~J.\ 2009.\ How Big Were the First Planetesimals? Does Size Matter?.\ 
Lunar and Planetary Institute Science Conference Abstracts 40, 1760. }

\item[$\bullet$] Wetherill, G.~W.\ 1989.
Origin of the asteroid belt.\ {\it Asteroids II} 661-680.

\item[$\bullet$] Wetherill,
G.~W., Stewart, G.~R.\ 1989. Accumulation of a swarm of small
planetesimals.\ {\it Icarus} 77, 330-357.

\item[$\bullet$] Wetherill, G.~W. 1992. An
alternative model for the formation of the asteroids.\ {\it Icarus} 100,
307-325.

\item[$\bullet$] Wetherill,
G.~W., Stewart, G.~R. 1993. Formation of planetary embryos - Effects of
fragmentation, low relative velocity, and independent variation of
eccentricity and inclination.\ {\it Icarus} 106, 190-205.

\item[$\bullet$] Wurm, G., Blum, J.\ 
1998.\ Experiments on Preplanetary Dust Aggregation.\ Icarus 132, 125-136. 

\item[$\bullet$] Wurm, G., Blum, J.\ 
2000.\ An Experimental Study on the Structure of Cosmic Dust Aggregates and 
Their Alignment by Motion Relative to Gas.\ Astrophysical Journal 529, 
L57-L60. 

\item[$\bullet$] Wurm, G., Paraskov, G., 
Krauss, O.\ 2005.\ Growth of planetesimals by impacts at $\sim$25 m/s.\ 
Icarus 178, 253-263. 

\item[$\bullet$] Youdin, A.~N., 
Goodman, J.\ 2005.\ Streaming Instabilities in Protoplanetary Disks.\ 
Astrophysical Journal 620, 459-469. 

\end{itemize}

\end{document}